\documentclass{aa}  

\usepackage{graphicx}
\usepackage{txfonts}
\usepackage[colorlinks=true,citecolor=blue]{hyperref}
\begin{document}

\title{Hubble Constant and Mass Determination of Centaurus A \& M83 from TRGB Distances}

\author{Adrian Faucher\inst{1,2}
        \and David Benisty\inst{3}
        \and David F. Mota\inst{1}
}

\institute{Institute of Theoretical Astrophysics, University of Oslo, 0315 Oslo, Norway, \and 
École polytechnique, Route de Saclay, 91128 Palaiseau, France\\
\email{adrian.faucher@polytechnique.edu}
\and
Leibniz-Institut für Astrophysik Potsdam, An der Sternwarte 16, D-14482 Potsdam, Germany\\
\email{dbenisty@aip.de}}

   \date{}

% \abstract{}{}{}{}{} 
% 5 {} token are mandatory
 
  \abstract
{ An independent determination of the Hubble constant is crucial in light of the persistent tension between early- and late-Universe measurements. In this study, we analyze the dynamics of the Centaurus A (CenA) and M83 galaxies, along with their associated dwarf companions identified via Tip of the Red Giant Branch (TRGB) distance measurements, to constrain both the group mass and the local value of $H_0$. By examining the motions of these galaxies relative to the system’s barycenter, we apply both the minor and major infall models, which provide bounds on the true radial velocity dispersion.  {From the overlap from these approaches, we obtain a virial mass estimate of $(7.3\pm 2.0)\times10^{12}\ M_{\odot}\;$ and a Hubble flow-based mass of $(2.6 \pm 1.4) \times 10^{12}\,M_{\odot}$.} Modeling the cold Hubble flow around the group center-of-mass, we derive a corresponding value of the Hubble constant as $\left(64.0 \pm 4.6\right)\,\mathrm{km\,s^{-1}\,Mpc^{-1}}$. These results offer an independent, dynamically motivated constraint on the local value of $H_0$, explicitly accounting for the impact of peculiar velocities in the nearby Universe. We also discuss the $\sim 2\sigma$ tension between the virial and Hubble flow-based mass estimates, which likely arises from the fact that M83 is close to the velocity surface and break the Hubble flow model assumptions.  {While the Hubble flow fit emphasizes galaxies that follow smooth expansion on the lower branch of the velocity-distance relation, the virial mass estimate is found to be in good agreement with the group mass derived from the K-band luminosity of its brightest members and the projected mass methods.}
}


   \maketitle
%
%-----------------------------------------------------------------

\section{Introduction}

A long-standing discrepancy persists between measurements of the Hubble constant in the early and late Universe: $H_0 = 67\,\mathrm{km\,s^{-1}\,Mpc^{-1}}$ as inferred from Planck CMB observations~\citep{Planck:2018vyg}, versus $H_0 = 73\,\mathrm{km\,s^{-1}\,Mpc^{-1}}$ derived from local Cepheid-supernova distance ladders~\citep{SupernovaCosmologyProject:1998vns}. This so-called ``Hubble tension'' has emerged as one of the most significant challenges in modern cosmology, potentially pointing to unknown systematics or a breakdown in the standard $\Lambda$CDM model~\citep{CosmoVerse:2025txj,Dainotti:2021pqg,DeSimone:2024lvy,Dainotti:2025qxz}. Resolving this tension is crucial, as the value of $H_0$ governs not only the current expansion rate of the Universe, but also its inferred age and ultimate fate. Independent determinations of $H_0$ based on local dynamics offer a complementary test of this discrepancy, free from many of the assumptions underpinning both early- and late-time probes. However, local measurements are complicated by peculiar velocities that arises from the gravitational interactions with nearby mass concentrations. To disentangle these effects from the cosmic expansion, one must account for both the peculiar velocity field and its cosmological context~\citep{Pavlidou:2013zha,Benisty:2023vbz,Benisty:2024tlv}.

The CenA (NGC 5128) and M83 (NGC 5236) galaxy complex provides a powerful laboratory for investigating local cosmic expansion dynamics, located at distances of $3.8 \pm 0.1$ Mpc and $\sim 4.7$ Mpc, respectively \citep{Karachentsev:2007AJ,Tully:2015AJ,Muller:2019}. Owing to its intermediate position between virialized structures and the onset of the linear Hubble flow, the CenA/M83 system serves as a valuable testbed for probing gravitational perturbations, infall kinematics, and velocity dispersion with minimized contamination from large-scale structure. Its well-studied constituents—especially the massive and active CenA enable precise barycentric modeling grounded in robust observational data.

The system is composed of two dynamically distinct subgroups: a dominant, massive CenA component with a virial mass of $[6.4-8.1] \times 10^{12}\, M_{\odot}$ \citep{Karachentsev:2007AJ}, and a less massive M83 subgroup $[1.3-3] \times 10^{12}\,M_{\odot}$ noted for intense star formation and evidence of tidal interactions \citep{Muller:2025}. Deep imaging has revealed 16 additional dwarf galaxy candidates in the M83 subgroup \citep{Muller:2015}, whose spatial alignment in a single dominant plane nearly perpendicular to the dust lane of CenA hints at coordinated infall and previous dynamical interactions \citep{Muller:2016}, an additional 41 new candidates were found in an expanded survey of 500 sq. deg around the Cen A and M83 subgroups by \cite{Muller:2017}.

\begin{figure*}[t!]
\centering
\includegraphics[width=0.75\textwidth]{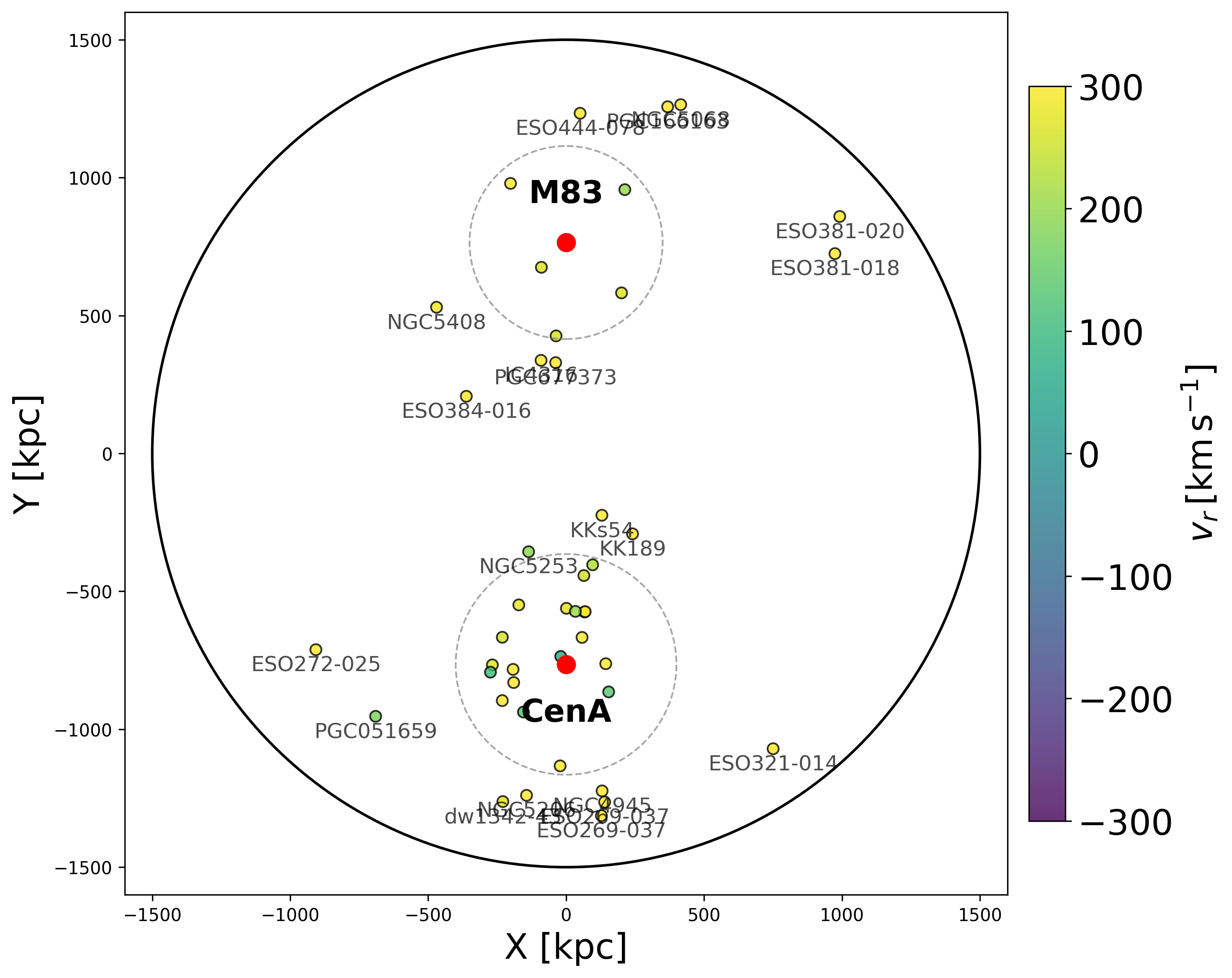}
\caption{Spatial distribution of galaxies in the CenA/M83 group shown in rotated Cartesian coordinates centered on the midpoint between Cen A and M83. Symbols are color–coded by heliocentric radial velocity (km\,s$^{-1}$). Dashed gray circles indicate the individual virial regions of Cen A and M83, and the black circle marks the 1.5 Mpc zero-velocity surface.}
\label{fig:skyplot}
\end{figure*}

The system as a whole seems to run away from us with a well-defined zero-velocity surface at $1.40 \pm 0.11$ Mpc \citep{Karachentsev:2007AJ}, indicating the gravitational boundary of the complex. HI observations further reveal extended tidal features and velocity bridges between the galaxies \citep{Muller:2021c,Muller:2022}, confirming ongoing interactions that influence their orbital evolution. Located on the outskirts of the dominant gravitational centers of the local volume, this research study the Hubble Flow around CenA/M83 by determining the barycenter of the system and the spread of the galaxies around the local Hubble Flow. Recent dynamical models reinforce its transitional nature, showing that flow perturbations decay within this region as the Universe approaches large-scale homogeneity \citep{Muller:2021,Muller:2024,Muller:2025}.

An important caveat concerns the kinematics of satellites in the immediate vicinity of Cen A, where most of the dwarf galaxies have been shown to reside in a co-rotating planar structure \citep{Muller:2018hks,Muller:2021}. Since our analysis focuses on galaxies in the surrounding Hubble flow, at distances of approximately [1.5–6] Mpc from the group center, the corotating satellite plane does not directly affect our modeling. This distinction is consistent with previous studies of the LG, where planar satellite distributions around the Milky Way and M31 are acknowledged, but the Hubble flow analysis relies on the more distant galaxy population \citep{Makarov:2025}. As also demonstrated in N-body simulations \citep{Penarrubia:2014oda}, the large-scale flow at these distances approaches spherical symmetry despite internal anisotropies within virialized regions. We therefore note that while coherent satellite planes are relevant within the virial radius, the present study addresses the dynamics of galaxies at larger scales, where the transition to the local cosmic expansion dominates.

In this paper, we determine the Hubble constant by analyzing the peculiar motion of this system. Using Tip of the Red Giant Branch (TRGB) distances calibrated via the latest distance ladder refinements combined with HI kinematics, we model their mutual gravitational interaction and infall trajectory relative to the LG (LG) frame. This approach isolates the peculiar velocity component, enabling direct comparison to the expected Hubble flow at their measured distance and yielding a high-fidelity $H_0$ estimate with minimized local flow contamination. Section~\ref{sec:psdets} describes the observational data sets and the determination and barycenter modeling that underpin our kinematic analysis. Section~\ref{sec:par} reports the derived constraints on $H_0$ and the total group mass, including a Bayesian treatment of uncertainties.  
Finally, Section~\ref{sec:Dis} discusses the implications of these results and summarizes our conclusions.

\section{Phase Space Determination}
\label{sec:psdets}
\subsection{The Data}  

In this study, we employ data from several extragalactic catalogs to analyze galaxies within the CenA/M83 galaxy group. Distance measurements are primarily adopted from the catalog Cosmicflows 4 (CF4)~\citep{Tully:2022rbj}, which provides TRGB distances independently of any assumption on the Hubble constant. Radial velocities are obtained from the LEDA database ~\cite{Makarov:2014} accessed via the Extragalactic Distance Database (EDD)~\citep{Tully:2009EDD}. Additionally, we incorporate velocities and TRGB distances from recent studies of dwarf galaxies in the Cen A and M83 groups \citep{Muller:2021,Muller:2025}, which also provide TRGB distances. Radial velocities obtained from LEDA are initially in the geocentric frame and are corrected to the LG frame~\citep{Karachentsev:1996AJ}. Distances across all datasets are based on TRGB determinations, ensuring consistency and independence from redshift-based methodologies. Consequently, we are able to accurately reconstruct the spatial and kinematic structure of the CenA/M83 group, free from biases introduced by cosmic expansion assumptions. Our sample selection includes only those galaxies that have both distance and radial-velocity measurements with associated uncertainties and that lie close to the inferred barycenter of the CenA/M83 system. As we later assume that the barycenter lies along the line connecting Cen A and M83, we initially selected all galaxies within 10 Mpc of either Cen A or M83. This yields a final sample of 135 galaxies from the EDD and 22 additional galaxies from recent studies. The selected galaxies span distances from $0.9$ Mpc to $14.7$ Mpc and have LG–frame velocities ranging from $-10$ km s$^{-1}$ to $1300$ km s$^{-1}$.

\begin{figure}[t!]
\centering
\includegraphics[width=0.28\textwidth]{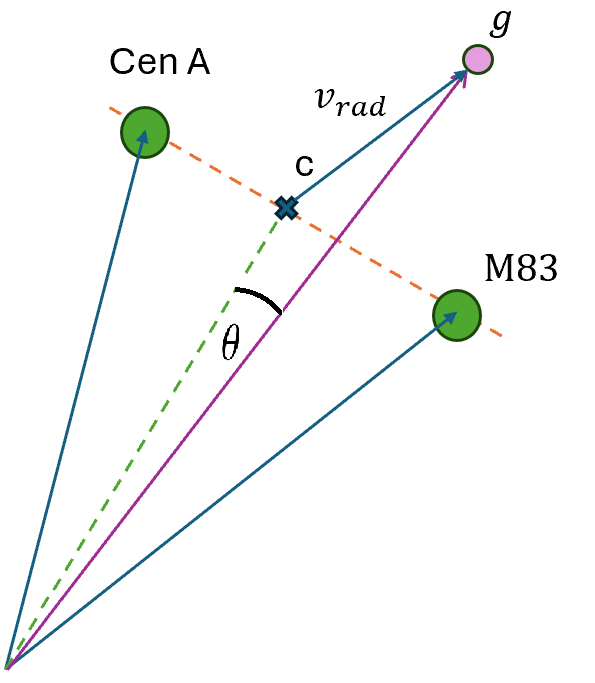}
\caption{Schematic illustration of barycenter ($c$) determination along the line connecting the CenA and M83 galaxies. For a given dwarf galaxy ($g$), the radial velocity is estimated using either the minor or major infall model. The barycenter is identified by minimizing the velocity dispersion of the radial velocities towards the barycenter. The angle $\theta$ is the angle between the angle between galaxy (g) and the center (c). }
\label{fig:com_det}
\end{figure}

\subsection{Barycenter determination}  
To analyze the dynamics of galaxy pairs within the CenA/M83 system, we transform observed velocities into the barycenter frame. Observational constraints limit velocity measurements to the Line-of-Sight (LoS) components. Consequently, we require dynamical models to estimate radial infall velocities relative to the CenA/M83 barycenter we adopt two limiting cases~\citep{bib:Karachentsev2006,Wagner:2025wrp}:
\begin{itemize}
    \item The \textbf{Minor infall model:} Assumes negligible transverse velocities ($v_{\perp,\mathrm{c}} = v_{\perp,j} = 0$). The radial infall velocity for galaxy $j$ relative to the Center of Mass (CoM) ($c$) is:  
\begin{align}  
v_{\mathrm{rad,min}} = \frac{v_{\mathrm{c}} {r}_{\mathrm{c}} + v_{j}  {r}_{j} - \cos \theta_{\mathrm{c},j}  ( v_j  {r}_{\mathrm{c}} + v_\mathrm{c} {r}_{j} )}{r_{gc}}.  
\label{eq:v_min}  
\end{align}  

This symmetric treatment minimizes peculiar velocity contributions, representing the most probable scenario when transverse motions are subdominant.

\item The \textbf{Major infall model:} Assumes zero transverse CoM velocity and negligible galaxy tangential velocity ($v_{\perp,c} = v_{\text{tan}} = 0$). The infall velocity becomes:  
\begin{align}  
v_{\mathrm{rad,maj}} = \frac{v_j - v_\mathrm{c} \cos \theta_{\mathrm{c},j}}{r_j - r_\mathrm{c} \cos \theta_{\mathrm{c},j}} r_{gc}.  
\label{eq:v_maj}  
\end{align}  
Note that for this model, galaxies located very close to the centre of mass can make the denominator arbitrarily small, resulting in non-physical velocity values.. In all subsequent analyses, these galaxies are excluded.
\\\\
As shown in \cite{Wagner:2025wrp,Benisty:2025upv}, statistically, the minor infall is underestimate the true radial velocity and the major infall is overestimate the radial velocity.

\end{itemize}

\begin{figure}[t!]
\centering
\includegraphics[width=0.45\textwidth]{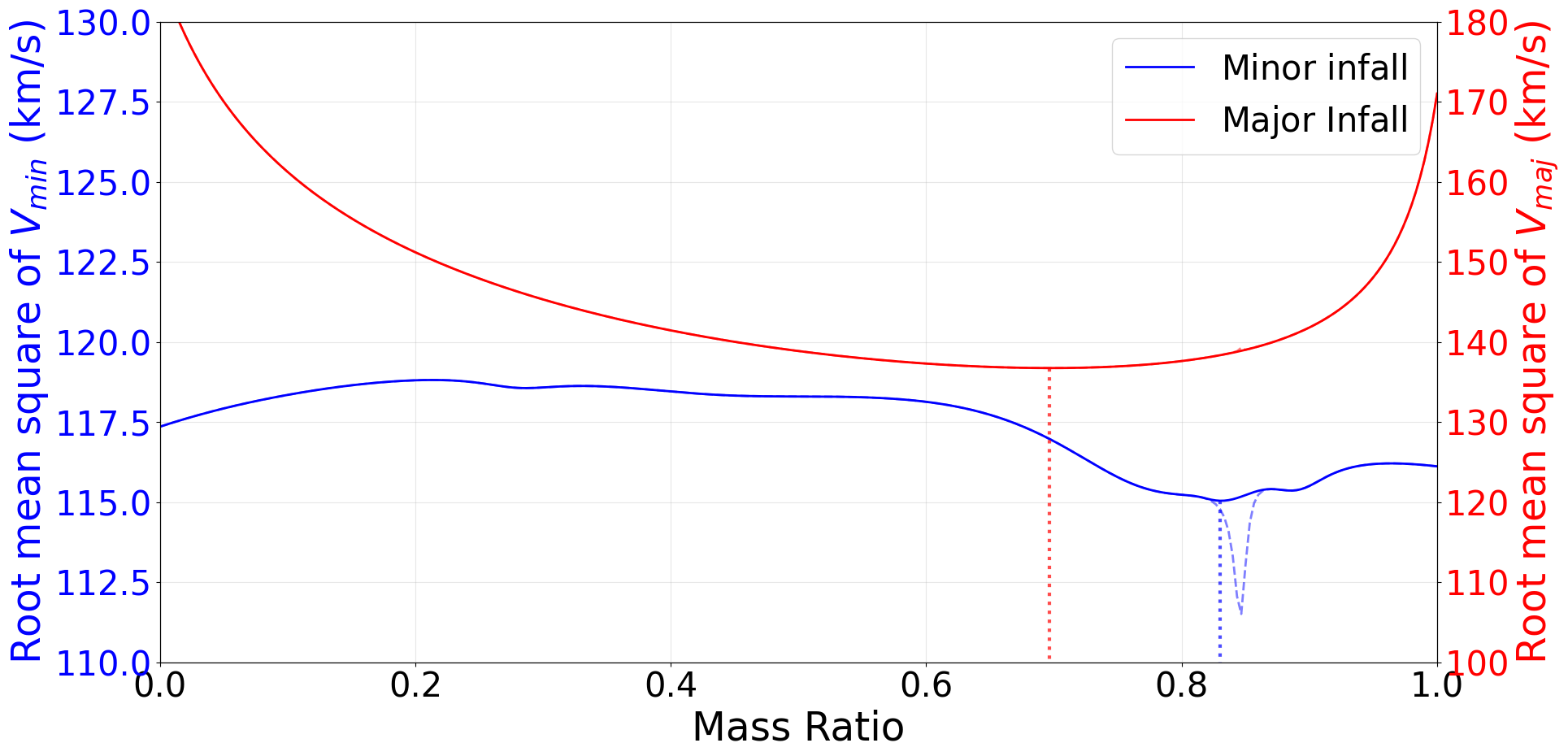}
\caption{ {Root mean square of the relative velocity dispersion, $\sqrt{\langle v^2 \rangle}$, with respect to the CenA/M83 system’s barycenter, shown as a function of the mass ratio $\bar{m}_{CenA} = m_{CenA} / m_{\mathrm{Tot}}$. Velocities are measured in the Local Group frame under the minor infall (blue) and major infall (red) models. Dotted lines: Raw results of $\sqrt{\langle v^2 \rangle}$ as a function of $\bar{m}{\mathrm{CenA}}$, mostly lying slightly below the smoothed (solid) curves, except for a local spike near $\bar{m}_{\mathrm{CenA}} \approx 0.85$ in the minor infall case. This spike is interpreted as a numerical bias arising from the clustering of galaxies near the CenA/M83 barycenter position (\citealp{Muller:2021,Muller:2025}). Vertical dashed lines: Positions of the minima in $\sqrt{\langle v^2 \rangle}$ for each model, indicating the best-fitting mass ratio. To avoid confusion, these vertical lines are plotted with a distinct dashed style, different from the dotted raw data lines.}}

\label{fig:com_velocity_dispersion}
\end{figure}

Understanding the motion of galaxies within the group requires careful modeling of their velocity field. The analysis begins by transforming the observed heliocentric velocities into a reference frame of LG~\citep{Karachentsev:1996AJ}. We assume that the dynamical center lies along the axis connecting the CenA and M83 halos with its exact position determined by their mass ratio. Similarly to Ref.~\citet{Karachentsev:2008st}, we parameterize the barycenter velocity $\vec{v}_{\mathrm{c}}$ using the mass ratio $\bar{m}_{CenA}$: 
\begin{equation}
\vec{v}_{\mathrm{c}} = \bar{m}_{CenA} \vec{v}_{\mathrm{CenA}} + (1 - \bar{m}_{CenA}) \vec{v}_{\mathrm{M83}},
\end{equation}
where $\bar{m}_{CenA}$ represents the fractional mass contribution of CenA relative to the total mass of the system, and then project this velocity on the barycenter line-of-sight vector to obtain the line of sight velocity. This formulation allows us to explore the kinematic structure across possible mass ratios: $\bar{m}_{CenA} = 1$ corresponds to a CenA-centered frame, while $\bar{m}_{CenA} = 0$ gives an M83-centered frame.  

\begin{figure*}[t!]
\centering
\includegraphics[width=0.75\textwidth]{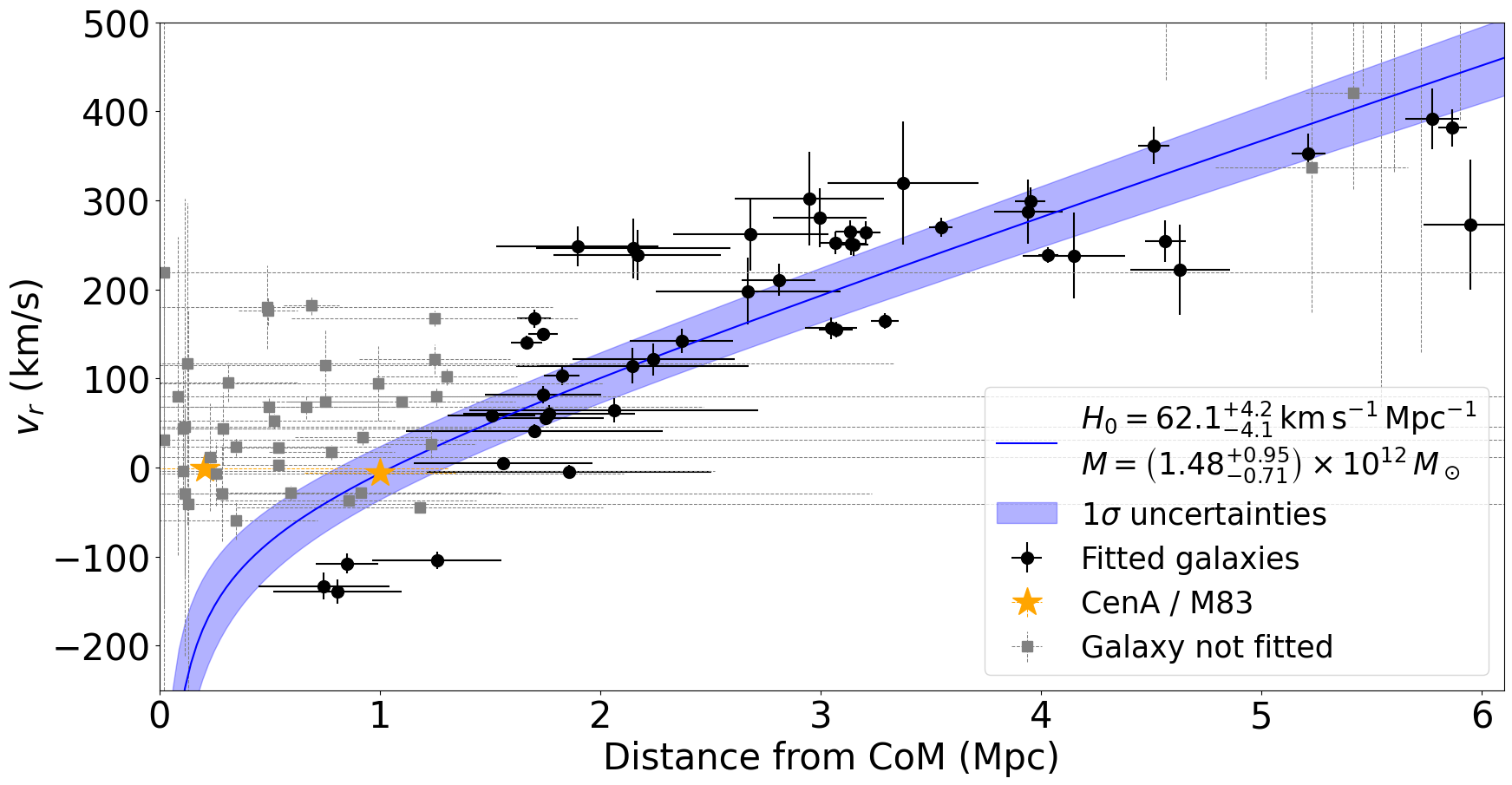}
\includegraphics[width=0.75\textwidth]{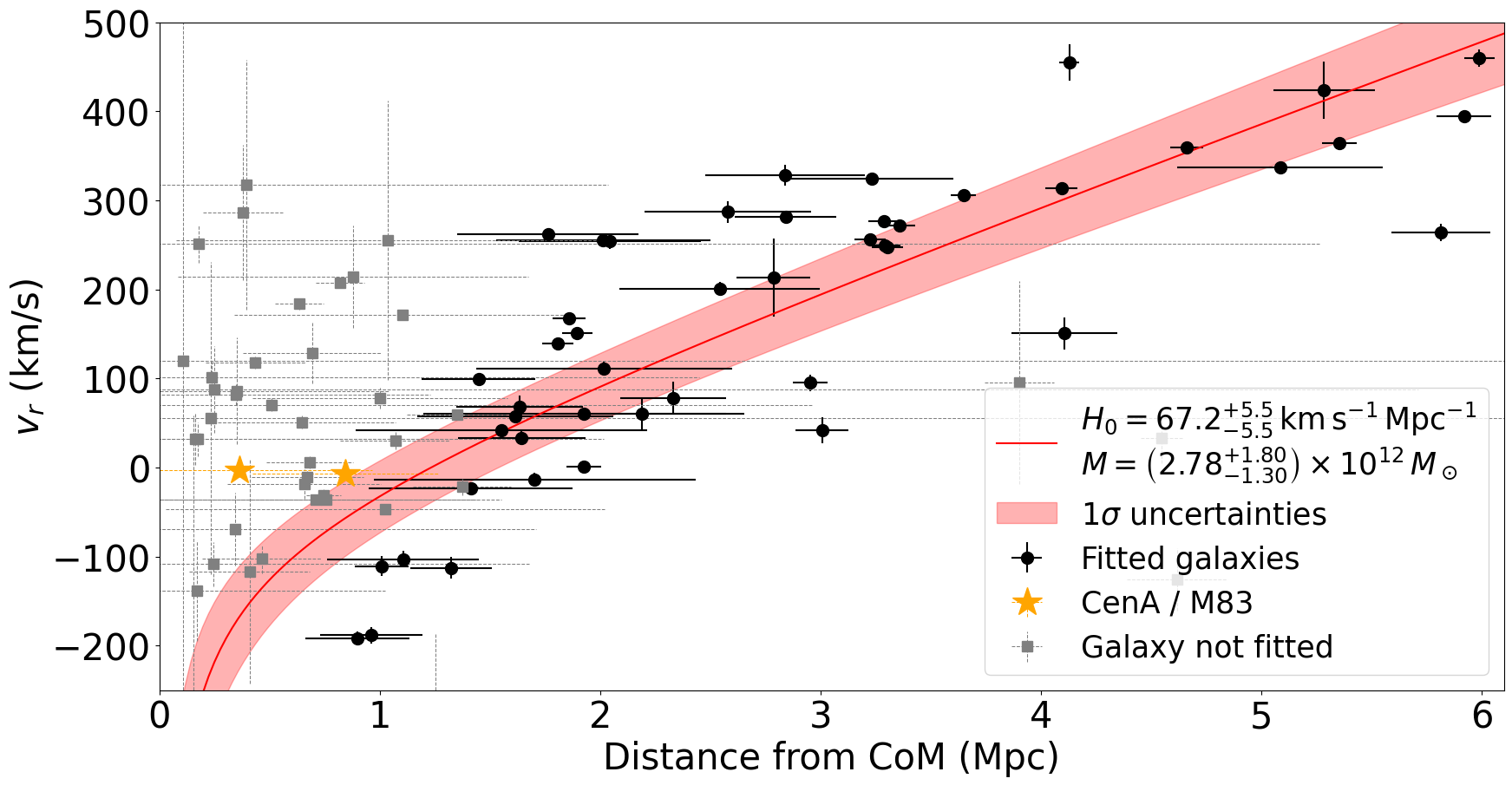}
\caption{The radial velocity vs. the 3D distance from the group barycenter for the minor (upper) and the major (lower) infall models. The Muse galaxies are shown in green and the galaxies that use for fit are shown in black. The Hubble Flow fit model is presented with $3\sigma$ error. }
\label{fig:post}
\end{figure*}
The determination of the barycenter position requires a fixed set galaxies with position and velocities that are always usable. so the galaxies used for this analysis are those which are closer than $3 Mpc$ of the barycenter and have $\lvert \Delta v_r \rvert < 500 \text{km.s}^{-1}$ for all possible barycenter (the second restriction excludes galaxies with non-physical velocities).

For each trial mass ratio, we compute minor and major infall velocities relative to the CoM and evaluate the velocity dispersion $\sqrt{\langle v^2 \rangle}$ for selected galaxies. The optimal $\bar{m}_1$ minimizes this dispersion, identifying the most stable dynamical frame. Fig.~(\ref{fig:com_velocity_dispersion}) shows $\sqrt{\langle v^2 \rangle}$ as a function of $\bar{m}_1$, revealing distinct minima for both infall models: the major infall model curve gives the minimum at $\bar{m}_{CenA} \approx 0.70$, which gives a distance from us of $r_c = 3.95 \text{Mpc}$ and a velocity from the LG CoM of $v_c = 302 \text{km.s}^{-1}$ and the minor infall model curve gives the minimum at $\bar{m}_{CenA} \approx 0.83$ which gives a distance from us of $r_c = 3.80 \text{Mpc}$ and a velocity from the LG CoM of $v_c = 304 \text{km/s}^{-1}$. This kinematic minimization provides a model-dependent mass ratio estimate independent of luminosity-based methods. The difference between models reflects systematic uncertainties in the assumptions of transverse velocity, which we propagate into our $H_0$ constraints. 

\section{Parameter Determination}
\label{sec:par}

\subsection{Hubble Flow Fit}
A fundamental tool for analyzing these systems is the velocity-distance relationship. \citet{bib:Sandage1986} introduced a method to estimate the mass of galaxy systems using this relationship, which encapsulates the kinematic state of the system. By measuring velocities and distances of member galaxies, one can estimate both the total mass of the system and the Hubble constant through nonlinear fitting techniques~\citep{Teerikorpi:2010zz}. This method has been applied to groups and clusters~\citep{Peirani:2005ti,Peirani:2008qs,DelPopolo:2022sev,Kim:2020gai,Penarrubia:2014oda,bib:Nasonova2011,Benisty:2025tct,Benisty:2025mfv,Wagner:2025wvh}. The concept of turnaround where gravitational attraction balances cosmic expansion—has been proposed as a stringent test of the $\Lambda$CDM cosmological model~\citep{Pavlidou:2013zha,Tanoglidis:2014lea,Pavlidou:2014aia,Pavlidou:2020afx,Paraskevas:2023itu}. The Hubble flow around the LG has been particularly informative in studying the dynamics of dwarf galaxies relative to cosmic expansion~\citep{Karachentsev:2008st}. We model the interplay between gravitational collapse and cosmic expansion through a nonlinear fit to CenA/M83 kinematics, via the relation from~\citet{Penarrubia:2014oda}:
\begin{equation}
v = \left( 1.2+0.31 \Omega_{\Lambda} \right) H_0 r -1.1 \sqrt{\frac{GM}{r}}
\end{equation}
with $\Omega_{\Lambda}=0.67$, our framework incorporates the zero-gravity radius demarcating bound and unbound regions \citep{Teerikorpi:2010zz}, dark energy effects via the cosmological constant \citep{Peirani:2008qs,DelPopolo:2022sev}, and turnaround radius tests of $\Lambda$CDM cosmology as \citet{Pavlidou:2013zha, Tanoglidis:2014lea} claim.

We select fitted galaxies as those lying between $ [1.40,6]\, $ Mpc from the center of mass, within the Hubble-flow velocity range.The inner limit of $1.40 \, \text{Mpc}$ corresponds to the 
zero-velocity radius from \citep{Karachentsev:2007AJ}; galaxies inside this radius are not expected 
to follow the Hubble law. The outer limit of $6 \, \text{Mpc}$ is imposed because beyond this scale 
other large-scale structures may significantly influence galaxy kinematics. For a better $M_{group}$ estimate, we also add few galaxies with $r<1.40 \text{Mpc}$, which fall towards the CoM ($ r < 1.40 \text{Mpc}$ and $v_r<-100 \text{km.s}^{-1}$).  Galaxies with uncertainty estimates smaller then 10\% on the velocity and 100\% on the distance are not used.
We fit the velocity-distance relation using a Bayesian approach. The likelihood of observing galaxies at distances $r_i$ from the CenA-M83 group CoM, with radial velocities $r_i$, given the parameter $H_0$, $M$ and the intrinsic velocity dispersion $\delta_I$ is:
\begin{equation}
     p(v|r,H_0,M,\delta_I) = -\sum_{i}{\frac{(f(r_i)-v_i)^2}{2\sigma_I^2}} + \ln{2\pi\sigma_I^2}
\end{equation}
with 
\begin{equation}
    \sigma_I= \text{err}(v_i)^2 + \text{err}(r_i)^2 \times (\frac{\partial f}{\partial r})^2 + \delta_I^2
\end{equation}
Where $\text{err}(v_i)$ and $\text{err}(r_i)$ are measurements errors and $\delta_I$ is the intrinsic velocity dispersion interpreted as scatter from peculiar motion.

\begin{figure}[t!]
\centering
\includegraphics[width=0.45\textwidth]{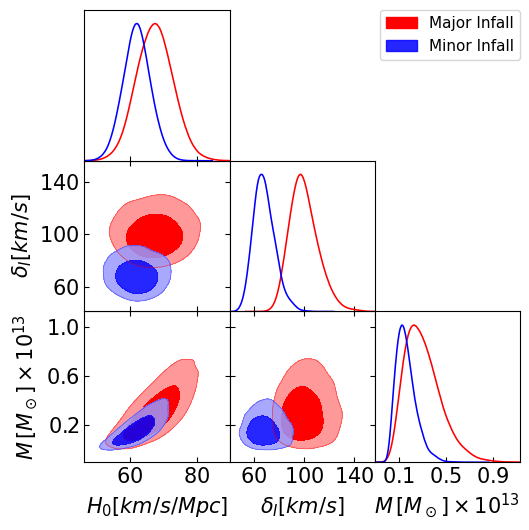}
\caption{Posterior distributions for the Hubble Constant $H_0$, the group mass $M\, [10^{12}\,\odot]$, and intrinsic velocity dispersion $\delta_I\,[km/s]$. The blue contour shows the minor infall model while the red shows major infall posterior.}
\label{fig:posterior_hubble}
\end{figure}

We employed the \textit{emcee} sampler \citep{Foreman:2013} to explore the posterior distributions of key model parameters, as illustrated in Fig.~(\ref{fig:posterior_hubble}). These include the Hubble constant $H_0$, the group mass $M_{\mathrm{group}}$, and the intrinsic velocity dispersion $\delta_I$. For the \textit{minor infall model}, the analysis yields $H_0 = 62.1^{+4.2}_{-4.1}~\mathrm{km\,s^{-1}\,Mpc^{-1}}$, $M_{\mathrm{group}} = 1.48^{+0.95}_{-0.71} \times 10^{12}~M_\odot$, and $\delta_I = 67.7^{+8.8}_{-7.8}~\mathrm{km\,s^{-1}}$. In contrast, the \textit{major infall model} gives a slightly higher Hubble constant, $H_0 = 67.2^{+5.5}_{-5.5}~\mathrm{km\,s^{-1}\,Mpc^{-1}}$, a larger group mass of $M_{\mathrm{group}} = 2.78^{+1.80}_{-1.30} \times 10^{12}~M_\odot$, and a higher intrinsic velocity dispersion of $\delta_I = 98.2^{+11.5}_{-9.7}~\mathrm{km\,s^{-1}}$.

These results highlight subtle but important differences between the two models. The minor infall model tends to \textit{underestimate} $H_0$, while the major infall model tends to \textit{overestimate} it. Moreover, radial trends shown in Fig.~\ref{fig:post} reveal three key features: $H_0$ estimates stabilize beyond $\sim1.5$~Mpc, where the Hubble flow dominates; mass estimates converge within $r < 0.8$~Mpc, where virial equilibrium is expected to hold; and the systematic offset in $H_0$ between models reflects differing assumptions about the nature of galaxy infall.  {The $2\sigma$ overleaping between the models give:}
\begin{equation}
H_0 = 64.0 \pm 4.6~\mathrm{km\,s^{-1}\,Mpc^{-1}} \, \& \,  M_{\rm Group} = (2.6 \pm 1.4)\times 10^{12}\,M_\odot,
\end{equation}
which we adopt as the best combined values accounting for the bias introduced by different infall assumptions.  However, the approach used to fit $H_0$ and $M_{\mathrm{group}}$ is not exact, due to the fact that the independent variable $r$ is itself subject to measurement error - a phenomenon known as regression dilution bias. A more accurate formulation of the likelihood function is given by:

\begin{align}
p(v|r,H_0,M,\delta_I) = -\sum_{i} \left[ \frac{(y_i - v_i)^2}{2(\delta_I^2 + \mathrm{err}(v_i)^2)} + \frac{(x_i - r_i)^2}{2\,\mathrm{err}(r_i)^2} \right] \nonumber \\ + \ln(2\pi \sigma_i^2), 
\end{align}
where $(x_i, y_i)$ is the orthogonal projection of the observed data point $(r_i, v_i)$ onto the model curve $y = f_{H_0, M}(x)$. Although the previous likelihood expression was similar to the standard form $\chi^2$ used in least-squares regression, the above formulation corresponds to the Deming regression method, which accounts for uncertainties in both variables. 

Implementing this likelihood in an MCMC sampler cannot directly fit for $\delta_I$. To approximate its effect, we artificially inflate the velocity uncertainties by adding the intrinsic dispersions obtained from the MCMC fit:  $\delta_I^{\mathrm{min}} = 68.1~\mathrm{km\,s^{-1}}$ for the minor infall model and $\delta_I^{\mathrm{maj}} = 98.4~\mathrm{km\,s^{-1}}$ for the major infall model. Using this approximation, we obtain $H_0 = 63.8 \pm 3.9~\mathrm{km\,s^{-1}\,Mpc^{-1}}$ and $M_{\mathrm{group}} = 1.81 \pm 0.84 \times 10^{12}~M_\odot$ for the minor infall case, and $H_0 = 68.1 \pm 5.2~\mathrm{km\,s^{-1}\,Mpc^{-1}}$ with $M_{\mathrm{group}} = 3.05 \pm 1.55 \times 10^{12}~M_\odot$ for the major infall case.

These results are consistent with those obtained from the complete posterior MCMC, confirming that the inclusion of $\delta_I$ dominates the uncertainty budget and effectively justifies the use of a least squares approximation as a proxy for orthogonal distance regression (ODR). A key advantage of the Bayesian framework is its capacity to incorporate observational uncertainties and prior information in a coherent, principled manner. However, assumptions regarding the functional form of intrinsic velocity dispersion and the symmetric treatment of distance and velocity errors may introduce subtle biases. Future work could explore hierarchical Bayesian models or Gaussian process regression to more flexibly account for these effects and better characterize the local velocity field.

\begin{figure}[t!]
\centering
\includegraphics[width=0.45\textwidth]{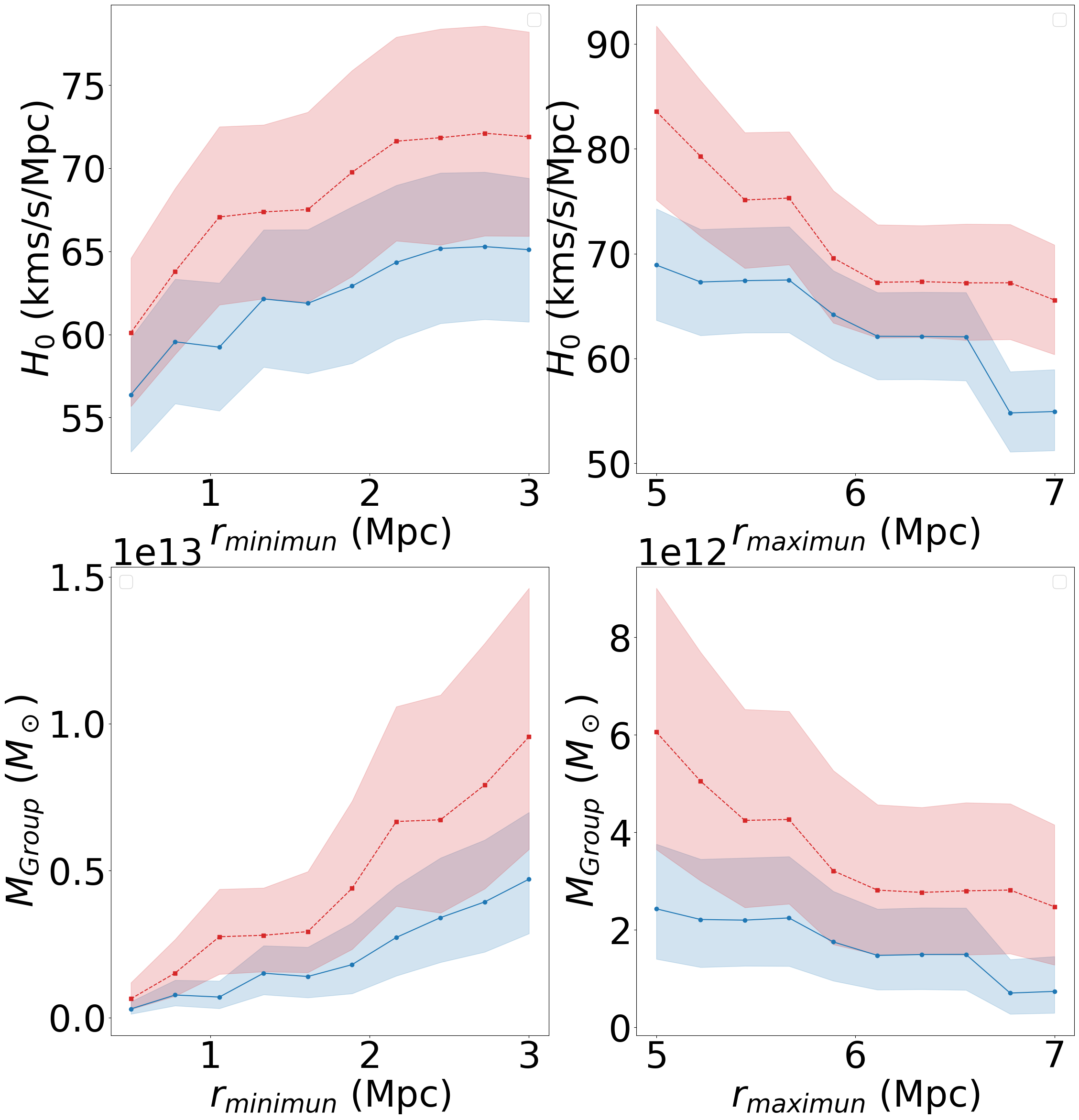}
\caption{Dependence of the fitted Hubble constant ($H_0$) and group mass ($M_{\mathrm{group}}$) on the choice of minimum and maximum selection radii. The blue curves correspond to the \textit{Minor Infall} model, and the red curves to the \textit{Major Infall} model. The colored area correspond to the 1 $\sigma$ uncertainty on the results \textbf{Top panels:} fitted values of $H_0$. 
\textbf{Bottom panels:} fitted values of $M_{\mathrm{group}}$. \textbf{Left:} variation of $H_0$ and $M_{\mathrm{group}}$ as a function of the minimum radius $r_{\mathrm{minimum}}$ used to selected galaxies for the fit. 
In these cases, the maximum radius was fixed at 6~Mpc, and galaxies close to the center of mass (CoM) and apparently infalling toward it were kept. \textbf{Right:} variation of $H_0$ and $M_{\mathrm{group}}$ as a function of the maximum radius $r_{\mathrm{maximum}}$. Here, the minimum radius was fixed at 1.4~Mpc, and nearby galaxies infalling toward the CoM were kept.}
\label{fig:min_max_fit}
\end{figure}

We also tested different values of the selection radii, with 
$r_{\mathrm{min}} \in [0.5~\mathrm{Mpc}, 3~\mathrm{Mpc}]$ 
and 
$r_{\mathrm{max}} \in [5~\mathrm{Mpc}, 7~\mathrm{Mpc}]$, 
which define the radial range of galaxies used for the Hubble law fitting. 
This allows us to assess the impact of galaxy selection on the fitted parameters $H_0$ and $M_{\mathrm{group}}$. 
The results are shown in Fig.~\ref{fig:min_max_fit}. 
When the minimum radius decreases below the zero-velocity surface, an increasing number of galaxies that do not follow the Hubble flow—because of the internal dynamics within the bound region—are included in the fit. Similarly, when the maximum radius exceeds 6~Mpc, more galaxies deviate from the Hubble law due to motions induced by other large-scale structures.

\begin{figure*}[t!]
\centering
\includegraphics[width=0.85\textwidth]{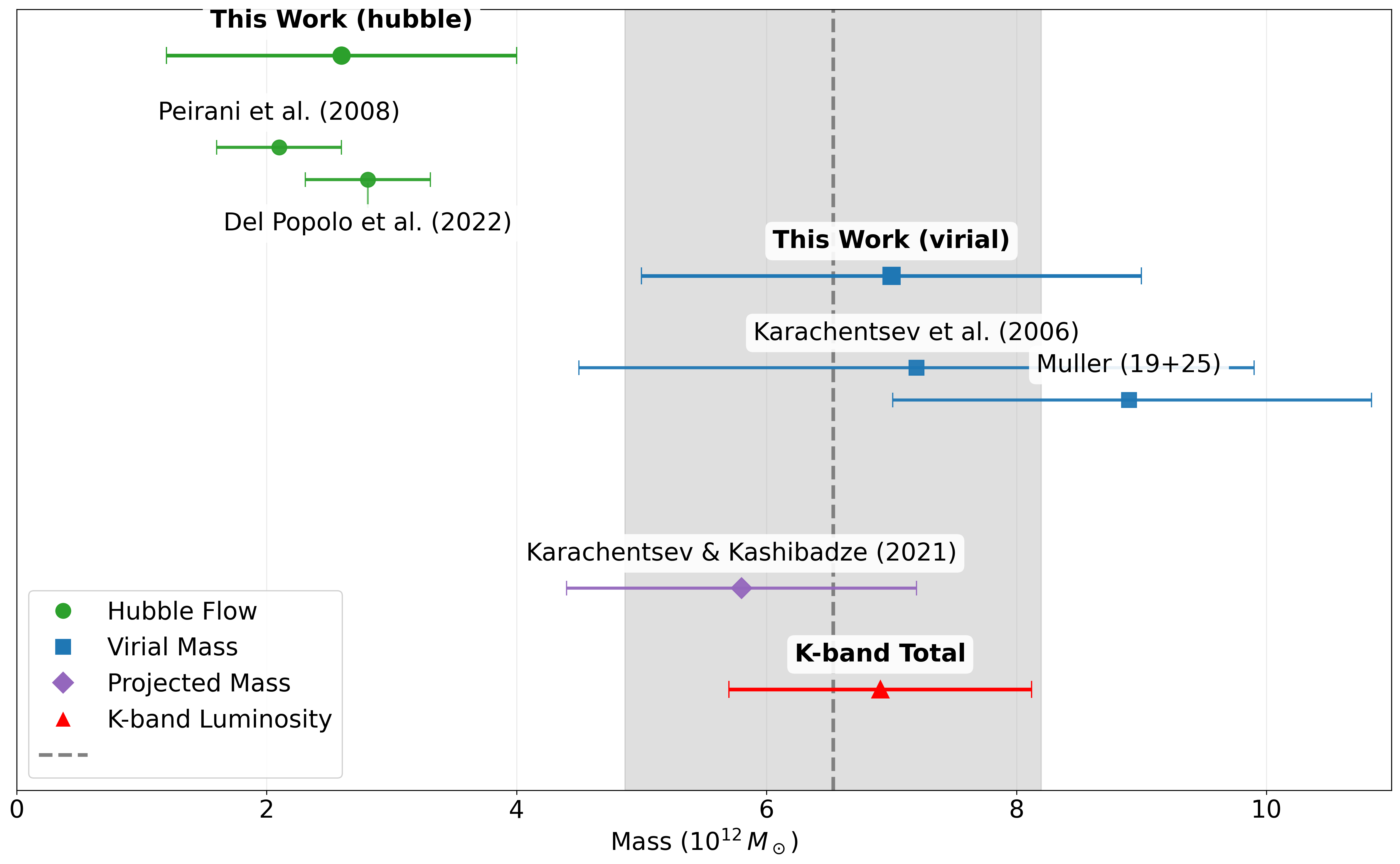}
\caption{ {Comparison of CenA/M83 galaxy group mass estimates obtained in this work with other recent estimates, taken from~\citet{Peirani:2008qs, Karachentsev:2007AJ, Muller:2019, Karachentsev:2021, DelPopolo:2021hkz, Muller:2025}. An overall estimate with $\left(6.53 \pm 1.66\right) \times 10^{12}\,M_{\odot}$ is added as a gray area.}}
\label{fig:final_mass}
\end{figure*}

\subsection{Virial Mass}
Assuming that the core of the CenA-M83 system encompasses galaxies within the velocity surface around the barycenter, we calculate the virial mass \citep{Limber:1960,bib:Bahcall1981,Heisler:1985,bib:Evans2011,Tully:2015AJ,Benisty:2024tlv}:
\begin{equation}  
M_{\text{vir}} = \alpha \frac{\sigma_\mathrm{v}^2 r_\mathrm{G}}{G} \;, \quad \text{where}  \quad r_{G0} = N/\sum{1/r_{i0}}  
\label{eq:virial_mass}  
\end{equation}
where $r_{G}$ runs over all of the dwarf galaxies towards the main halo. In the case of two main halos (CenA and M83), the harmonic radius is modified to be: $\frac{1}{r_{G}} = \frac{m_1}{r_{G,CenA}} + \frac{ 1 - m_1}{r_{G,M83}}$. $\sigma_\mathrm{v}$ represents the velocity dispersion of galaxies in the core, $r_\mathrm{G}$ denotes the harmonic mean of 3D pairwise distances between the $N$ galaxies (weighted by their inverse separations $r_{ij}$), and $\alpha$ is a dimensionless geometrical factor that depends on the dark matter distribution and velocity anisotropy.

For reference, $\alpha = 3$ corresponds to isotropic orbits in an isothermal sphere, $\alpha = 2.6$ to isotropic velocities in an NFW profile, and $\alpha = 2.4$ to specific anisotropy models~\citep{Limber:1960,bib:Bahcall1981}. To encompass uncertainties in the dark matter distribution, we adopt a broad uniform prior of $\alpha \in [2, 4]$. We adapt the area of up to the velocity surface. We use a gaussian prior on the velocities and the distances from us. The velocity dispersions for galaxies in the virial core is: $103 \pm 3~\text{km/s}$. Substituting these results into Eq.~\ref{eq:virial_mass}, we give the overall mass as
\begin{equation}
M_{\rm Vir}= \left(7.3 \pm 2.0\right) \times 10^{12} M_{\odot}\;.
\end{equation}

%Note that those velocity dispersions are slightly different from the ones obtained with the barycenter because of the different of methods used to selected the galaxies. As demonstrated in \cite{Wagner:2025wrp,Benisty:2025upv}, the major infall model overestimates the true radial velocity dispersion while the minor infall model underestimates it, indicating that the true dispersion lies between these values. Based on the ratios from the dispersions as reported in appendix~\ref{sec:appenA}, the true radial velocity dispersion is estimated to be: $\sigma_{\text{min}} = 141\pm 10 ~\text{km/s} \;$. Substituting these results into Eq.~\ref{eq:virial_mass}, giving the overall mass to be:
%
%\begin{equation}  
%M_{\text{vir}}^{\text{min}} = 6.41^{+3.42}_{-2.40} \times 10^{12} \,M_{\odot} \;, \quad  M_{\text{vir}}^{\text{maj}} = 14.7^{+7.99}_{-5.61} \times 10^{12} \,M_{\odot}. 
%\end{equation}
%
%The minor and major infall models bracket the true radial velocity dispersion, implying the virial mass lies between these estimates with an overall values of $M_{\rm vir} = 8.5^{+3.4}_{-2.0} \times 10^{12}\,M_\odot$. 

\subsection{K band luminosity mass}

\begin{table}[t!]
\centering

\label{tab:luminosity_data}
\begin{tabular}{|l|c|c|}
\hline\hline
Group & $\log(L_K / L_\odot)$ & $L_K$ $(L_\odot)$ \\
\hline\hline
CenA & 10.89 & $7.76 \times 10^{10}$ \\
M83 & 10.86 & $7.24 \times 10^{10}$ \\
\hline\hline
\end{tabular}

\caption{$K$-band Luminosity Data from \citet{Karachentsev:2021}.}

\end{table}
 {We can use the luminosity of the main galaxies to infer the total mass of the group. Following \citet{Tully:2015opa}, we estimate group masses using the empirical $K$-band luminosity-to-mass relation calibrated against the virial masses of groups, which provides a robust total mass estimate independent of stellar initial mass function assumptions. The relation reads:}
\begin{equation}
M_{\rm group} = 43\,h \left(\frac{L_K}{10^{10}L_\odot}\right)^{1.15} \times 10^{10}\,M_\odot,
\end{equation}
where $L_K$ is the luminosity of 2MASS $K_s$.
This relation implies a mass-to-light ratio
\begin{equation}
\frac{M}{L_K} = 43\,\left(\frac{L_K}{10^{10}L_\odot}\right)^{0.15}\,M_\odot/L_\odot,\end{equation}
which is fixed at $M/L_K = 43,M_\odot/L_\odot$ for $L_K \le 10^{10}\,L_\odot$ and at $M/L_K = 121\,M_\odot/L_\odot$ for $L_K \ge 10^{13}\,L_\odot$. Between these luminosity limits, the power-law captures the trend observed in nearby groups from numerical-action modeling and other studies. Systems dominated by elliptical galaxies carry roughly 25\% more mass per unit $K_s$ light than spiral-dominated systems, and the overall uncertainty in the luminosity-to-mass conversion is about 20\% ~\citep{Tully:2015opa}. Adopting $h = 0.7 \pm 0.1$ and the 2MASS luminosities $L_{K,{\rm CenA}} = 7.8 \times 10^{10},L_\odot$ and $L_{K,{\rm M83}} = 7.2 \times 10^{10},L_\odot$ from \citet{Karachentsev:2021} (summarized in table~\ref{tab:luminosity_data}), we find $M_{\rm CenA} = \left(3.97 \pm 0.98\right) \times 10^{12}\,M_\odot$ and $M_{\rm M83} = \left(2.93 \pm 0.72\right) \times 10^{12}\,M_\odot$,
where the CenA mass includes the 25\% elliptical correction. The total mass of the combined system is:
\begin{equation}
M_{\rm K-band} = \left( 6.91 \pm 1.21 \right)  \times 10^{12}\,M_\odot.
\end{equation}
These estimates assume the group mass prescription is valid in the luminosity interval considered. The quoted uncertainties reflect contributions from both the adopted value of $h$ and the intrinsic 20\% scatter in the luminosity-to-mass relation, but do not include other potential sources of error such as distance uncertainties, internal extinction corrections, dynamical disturbances, or variations in the stellar initial mass function.

\section{Discussion}
\label{sec:Dis}

Our analysis provides an independent, dynamical estimate of the Hubble constant and total group mass of the CenA/M83 system, using both the minor and major infall models to bracket the impact of unknown transverse motions. The values we derive for the Hubble constant range from $H_0 = 64 \pm 4$~km\,s$^{-1}$\,Mpc$^{-1}$ (minor infall) to $H_0 = 68 \pm 5$~km\,s$^{-1}$\,Mpc$^{-1}$ (major infall), with a total value of $H_0 = 68\pm 5$~km\,s$^{-1}$\,Mpc$^{-1}$. These values are consistent with, but systematically lower than, recent late-Universe measurements based on Cepheid-calibrated supernovae (e.g., $H_0 \approx 73$~km\,s$^{-1}$\,Mpc$^{-1}$), and are closer to early-Universe determinations from the CMB (e.g., $H_0 \approx 67.4$~km\,s$^{-1}$\,Mpc$^{-1}$ from \textit{Planck}~\cite{Planck:2018vyg}). These findings highlight the importance of local dynamical effects in reconciling discrepancies in $H_0$ measurements. The significant peculiar velocities observed in the Local Volume complicate direct extrapolation of redshifts to infer the Hubble flow, especially within $\sim 5$~Mpc.

The spread between the two infall models illustrates the systematic uncertainty arising from the unknown tangential velocities. The minor infall model assumes purely radial motion and likely underestimates the total infall velocity, thus producing a lower Hubble constant and smaller group mass. In contrast, the major infall model assumes more concentrated motion along the line of sight and provides an upper bound. Our results suggest that the true values likely lie between these extremes. The posterior distributions show consistent correlations between Hubble constant, group mass, and intrinsic dispersion, underscoring how LG dynamics imprint on the observed Hubble flow.

A recent analysis based on the CF4 catalog \citep{Tully:2022rbj,Duangchan:2025uzj} infers $H_{0}=74.6\pm0.8\ \mathrm{km\,s^{-1}\,Mpc^{-1}}$, a value higher than ours. This offset can be traced to several factors.  First, the CF4 distance scale ties together multiple indicators (Tully–Fisher, Fundamental Plane, surface-brightness fluctuations, and others) and is calibrated assuming a particular adopted value of the Hubble constant, whereas our distance ladder uses TRGB distances that are independent of $H_{0}$. Second, the treatment of peculiar velocities and bulk flows differs: CF4 analyses and ours weight nearby objects differently, and because peculiar motions constitute a larger fraction of the observed redshift at small distances, the inferred expansion rate is especially sensitive to flow modeling and sky coverage.  Third, selection and sample-correction procedures differ between the studies. Because peculiar-velocity corrections directly affect the measured dispersion and thus the derived expansion rate, analyses that rely more heavily on the very local volume are particularly vulnerable to under- or over-estimates arising from flow modeling.

In addition to the Hubble constant, we derive two independent estimates of the total mass of the CenA/M83 system. From the Hubble flow model we obtain $M_{Hubble} = \left(2.6 \pm 1.4 \right)\times10^{12},M\odot$. This value agree with earlier CenA estimates but are notably lower than our virial result of $M_{\rm vir} = (7\pm 2)\times10^{12},M\odot$, which includes the full velocity dispersion of galaxies within the turnaround radius.  {A tension therefore exists between the virial and the K-band to and Hubble–flow determinations (see Fig.~\ref{fig:final_mass}, which compares our virial estimate with previous virial results~\citep{Muller:2019,Muller:2025}, Hubble–flow fits~\citep{DelPopolo:2021hkz,DelPopolo:2022sev}, and K-band luminosity estimates~\citep{Tully:2015opa} and projected mass~\citep{Karachentsev:2021}). An overall estimate of the mass without the lower Hubble flow gives $\left(6.53 \pm 1.66\right) \times 10^{12}\,M_{\odot}$. }

The lower Hubble–flow mass is expected because the Hubble Flow mass method assumes a single concentrated mass and treats each satellite as a test particle that has moved outward with zero peculiar velocity since the Big Bang. Such an approximation is valid only when satellites lie far enough from the main halo to feel its gravity as if it were a point mass.
For example, in the Milky Way-M31 system, the satellites are sufficiently distant that a clean, cold Hubble flow is observed, and the method performs well~\citep{Karachentsev:2008st,Makarov:2025}.

In contrast, for CenA/M83 the situation is different: as Fig.~\ref{fig:post} shows, M83 lies close to the system’s velocity surface, so the assumption of a single, isolated point mass is violated.
The proximity of M83 and its interaction with CenA introduce internal dynamics that the simple Hubble–flow model cannot capture, which naturally results in a lower total mass. Therefore, we place greater weight on the virial estimate, which more fully reflects the true dynamical state of the system.

\begin{acknowledgements}
We thank Igor Karachentsev for useful discussions and comments. DB is supported by a Minerva Fellowship of the Minerva Stiftung Gesellschaft für die Forschung mbH. DFM thanks the Research Council of Norway for their support and the resources provided by UNINETT Sigma2 -- the National Infrastructure for High-Performance Computing and Data Storage in Norway. This article is based on work from the COST Action CA21136 - "Addressing observational tensions in cosmology with systematics and fundamental physics" (CosmoVerse) and the Cost Action CA23130 - “Bridging high and low energies in search of quantum gravity (BridgeQG), supported by COST (European Cooperation in Science and Technology). 
\end{acknowledgements}

\bibliographystyle{aa}
\bibliography{ref.bib}

\begin{thebibliography}{54}
\expandafter\ifx\csname natexlab\endcsname\relax\def\natexlab#1{#1}\fi

\bibitem[{Aghanim {et~al.}(2020)}]{Planck:2018vyg}
Aghanim, N. {et~al.} 2020, Astron. Astrophys., 641, A6, [Erratum: Astron.Astrophys. 652, C4 (2021)]

\bibitem[{{An} \& {Evans}(2011)}]{bib:Evans2011}
{An}, J.~H. \& {Evans}, N.~W. 2011, \mnras, 413, 1744

\bibitem[{{Bahcall} \& {Tremaine}(1981)}]{bib:Bahcall1981}
{Bahcall}, J.~N. \& {Tremaine}, S. 1981, \apj, 244, 805

\bibitem[{Benisty(2025)}]{Benisty:2025mfv}
Benisty, D. 2025 [\eprint[arXiv]{2504.03009}]

\bibitem[{Benisty {et~al.}(2024)Benisty, Chaichian, \& Tureanu}]{Benisty:2024tlv}
Benisty, D., Chaichian, M.~M., \& Tureanu, A. 2024, Phys. Lett. B, 858, 139033

\bibitem[{Benisty {et~al.}(2023)Benisty, Davis, \& Evans}]{Benisty:2023vbz}
Benisty, D., Davis, A.-C., \& Evans, N.~W. 2023, Astrophys. J. Lett., 953, L2

\bibitem[{Benisty \& Mota(2025)}]{Benisty:2025upv}
Benisty, D. \& Mota, D. 2025, Astron. Astrophys., 698, A43

\bibitem[{Benisty {et~al.}(2025)Benisty, Wagner, Haridasu, \& Salucci}]{Benisty:2025tct}
Benisty, D., Wagner, J., Haridasu, S., \& Salucci, P. 2025 [\eprint[arXiv]{2504.04135}]

\bibitem[{Dainotti {et~al.}(2021)Dainotti, De~Simone, Schiavone, Montani, Rinaldi, \& Lambiase}]{Dainotti:2021pqg}
Dainotti, M.~G., De~Simone, B., Schiavone, T., {et~al.} 2021, Astrophys. J., 912, 150

\bibitem[{Dainotti {et~al.}(2025)}]{Dainotti:2025qxz}
Dainotti, M.~G. {et~al.} 2025, JHEAp, 48, 100405

\bibitem[{De~Simone {et~al.}(2025)De~Simone, van Putten, Dainotti, \& Lambiase}]{DeSimone:2024lvy}
De~Simone, B., van Putten, M. H. P.~M., Dainotti, M.~G., \& Lambiase, G. 2025, JHEAp, 45, 290

\bibitem[{Del~Popolo \& Chan(2022)}]{DelPopolo:2022sev}
Del~Popolo, A. \& Chan, M.~H. 2022, Astrophys. J., 926, 156

\bibitem[{Del~Popolo {et~al.}(2021)Del~Popolo, Deliyergiyev, \& Chan}]{DelPopolo:2021hkz}
Del~Popolo, A., Deliyergiyev, M., \& Chan, M.~H. 2021, Phys. Dark Univ., 31, 100780

\bibitem[{Di~Valentino {et~al.}(2025)}]{CosmoVerse:2025txj}
Di~Valentino, E. {et~al.} 2025 [\eprint[arXiv]{2504.01669}]

\bibitem[{Duangchan {et~al.}(2025)Duangchan, Valade, Libeskind, \& Steinmetz}]{Duangchan:2025uzj}
Duangchan, C., Valade, A., Libeskind, N.~I., \& Steinmetz, M. 2025 [\eprint[arXiv]{2507.22236}]

\bibitem[{{Foreman-Mackey} {et~al.}(2013){Foreman-Mackey}, {Hogg}, {Lang}, \& {Goodman}}]{Foreman:2013}
{Foreman-Mackey}, D., {Hogg}, D.~W., {Lang}, D., \& {Goodman}, J. 2013, \pasp, 125, 306

\bibitem[{{Heisler} {et~al.}(1985){Heisler}, {Tremaine}, \& {Bahcall}}]{Heisler:1985}
{Heisler}, J., {Tremaine}, S., \& {Bahcall}, J.~N. 1985, \apj, 298, 8

\bibitem[{{Karachentsev} \& {Kashibadze}(2021)}]{Karachentsev:2021}
{Karachentsev}, I. \& {Kashibadze}, O. 2021, Astronomische Nachrichten, 342, 999

\bibitem[{{Karachentsev} \& {Kashibadze}(2006)}]{bib:Karachentsev2006}
{Karachentsev}, I.~D. \& {Kashibadze}, O.~G. 2006, Astrophysics, 49, 3

\bibitem[{Karachentsev {et~al.}(2009)Karachentsev, Kashibadze, Makarov, \& Tully}]{Karachentsev:2008st}
Karachentsev, I.~D., Kashibadze, O.~G., Makarov, D.~I., \& Tully, R.~B. 2009, Mon. Not. Roy. Astron. Soc., 393, 1265

\bibitem[{{Karachentsev} \& {Makarov}(1996)}]{Karachentsev:1996AJ}
{Karachentsev}, I.~D. \& {Makarov}, D.~A. 1996, \aj, 111, 794

\bibitem[{{Karachentsev} {et~al.}(2007){Karachentsev}, {Tully}, {Dolphin}, {Sharina}, {Makarova}, {Makarov}, {Sakai}, {Shaya}, {Kashibadze}, {Karachentseva}, \& {Rizzi}}]{Karachentsev:2007AJ}
{Karachentsev}, I.~D., {Tully}, R.~B., {Dolphin}, A., {et~al.} 2007, \aj, 133, 504

\bibitem[{Kim {et~al.}(2020)Kim, Kang, Lee, \& Jang}]{Kim:2020gai}
Kim, Y.~J., Kang, J., Lee, M.~G., \& Jang, I.~S. 2020, Astrophys. J., 905, 104

\bibitem[{{Limber} \& {Mathews}(1960)}]{Limber:1960}
{Limber}, D.~N. \& {Mathews}, W.~G. 1960, \apj, 132, 286

\bibitem[{{Makarov} {et~al.}(2025){Makarov}, {Makarov}, {Makarova}, \& {Libeskind}}]{Makarov:2025}
{Makarov}, D., {Makarov}, D., {Makarova}, L., \& {Libeskind}, N. 2025, \aap, 698, A178

\bibitem[{{Makarov} {et~al.}(2014){Makarov}, {Prugniel}, {Terekhova}, {Courtois}, \& {Vauglin}}]{Makarov:2014}
{Makarov}, D., {Prugniel}, P., {Terekhova}, N., {Courtois}, H., \& {Vauglin}, I. 2014, \aap, 570, A13

\bibitem[{{M{\"u}ller} {et~al.}(2021){M{\"u}ller}, {Fahrion}, {Rejkuba}, {Hilker}, {Lelli}, {Lutz}, {Pawlowski}, {Coccato}, {Anand}, \& {Jerjen}}]{Muller:2021c}
{M{\"u}ller}, O., {Fahrion}, K., {Rejkuba}, M., {et~al.} 2021, \aap, 645, A92

\bibitem[{{M{\"u}ller} {et~al.}(2015){M{\"u}ller}, {Jerjen}, \& {Binggeli}}]{Muller:2015}
{M{\"u}ller}, O., {Jerjen}, H., \& {Binggeli}, B. 2015, \aap, 583, A79

\bibitem[{M{\"u}ller {et~al.}(2017)M{\"u}ller, Jerjen, \& Binggeli}]{Muller:2017}
M{\"u}ller, O., Jerjen, H., \& Binggeli, B. 2017, Astron. Astrophys., 597, A7

\bibitem[{{M{\"u}ller} {et~al.}(2016){M{\"u}ller}, {Jerjen}, {Pawlowski}, \& {Binggeli}}]{Muller:2016}
{M{\"u}ller}, O., {Jerjen}, H., {Pawlowski}, M.~S., \& {Binggeli}, B. 2016, \aap, 595, A119

\bibitem[{{M{\"u}ller} {et~al.}(2022){M{\"u}ller}, {Lelli}, {Famaey}, {Pawlowski}, {Fahrion}, {Rejkuba}, {Hilker}, \& {Jerjen}}]{Muller:2022}
{M{\"u}ller}, O., {Lelli}, F., {Famaey}, B., {et~al.} 2022, \aap, 662, A57

\bibitem[{M{\"u}ller {et~al.}(2018)M{\"u}ller, Pawlowski, Jerjen, \& Lelli}]{Muller:2018hks}
M{\"u}ller, O., Pawlowski, M.~S., Jerjen, H., \& Lelli, F. 2018, Science, 359, 534

\bibitem[{M{\"u}ller {et~al.}(2021)M{\"u}ller, Pawlowski, Lelli, Fahrion, Rejkuba, Hilker, Kanehisa, Libeskind, \& Jerjen}]{Muller:2021}
M{\"u}ller, O., Pawlowski, M.~S., Lelli, F., {et~al.} 2021, Astron. Astrophys., 645, L5

\bibitem[{{M{\"u}ller} {et~al.}(2024){M{\"u}ller}, {Pawlowski}, {Revaz}, {Venhola}, {Rejkuba}, {Hilker}, \& {Lutz}}]{Muller:2024}
{M{\"u}ller}, O., {Pawlowski}, M.~S., {Revaz}, Y., {et~al.} 2024, \aap, 684, L6

\bibitem[{{M{\"u}ller} {et~al.}(2025){M{\"u}ller}, {Rejkuba}, {Fahrion}, {Pawlowski}, {Famaey}, {Libeskind}, {Heesters}, {Lelli}, {Hilker}, {Taibi}, \& {Pearson}}]{Muller:2025}
{M{\"u}ller}, O., {Rejkuba}, M., {Fahrion}, K., {et~al.} 2025, arXiv e-prints, arXiv:2504.20765

\bibitem[{{M{\"u}ller} {et~al.}(2019){M{\"u}ller}, {Rejkuba}, {Pawlowski}, {Ibata}, {Lelli}, {Hilker}, \& {Jerjen}}]{Muller:2019}
{M{\"u}ller}, O., {Rejkuba}, M., {Pawlowski}, M.~S., {et~al.} 2019, \aap, 629, A18

\bibitem[{{Nasonova} {et~al.}(2011){Nasonova}, {de Freitas Pacheco}, \& {Karachentsev}}]{bib:Nasonova2011}
{Nasonova}, O.~G., {de Freitas Pacheco}, J.~A., \& {Karachentsev}, I.~D. 2011, \aap, 532, A104

\bibitem[{Paraskevas \& Perivolaropoulos(2024)}]{Paraskevas:2023itu}
Paraskevas, E.~A. \& Perivolaropoulos, L. 2024, Mon. Not. Roy. Astron. Soc., 531, 1021

\bibitem[{Pavlidou {et~al.}(2020)Pavlidou, Korkidis, Tomaras, \& Tanoglidis}]{Pavlidou:2020afx}
Pavlidou, V., Korkidis, G., Tomaras, T., \& Tanoglidis, D. 2020, Astron. Astrophys., 638, L8

\bibitem[{Pavlidou {et~al.}(2014)Pavlidou, Tetradis, \& Tomaras}]{Pavlidou:2014aia}
Pavlidou, V., Tetradis, N., \& Tomaras, T.~N. 2014, JCAP, 05, 017

\bibitem[{Pavlidou \& Tomaras(2014)}]{Pavlidou:2013zha}
Pavlidou, V. \& Tomaras, T.~N. 2014, JCAP, 09, 020

\bibitem[{Pe\~narrubia {et~al.}(2014)Pe\~narrubia, Ma, Walker, \& McConnachie}]{Penarrubia:2014oda}
Pe\~narrubia, J., Ma, Y.-Z., Walker, M.~G., \& McConnachie, A. 2014, Mon. Not. Roy. Astron. Soc., 443, 2204

\bibitem[{Peirani \& de~Freitas~Pacheco(2006)}]{Peirani:2005ti}
Peirani, S. \& de~Freitas~Pacheco, J.~A. 2006, New Astron., 11, 325

\bibitem[{Peirani \& Pacheco(2008)}]{Peirani:2008qs}
Peirani, S. \& Pacheco, J. A. D.~F. 2008, Astron. Astrophys., 488, 845

\bibitem[{Perlmutter {et~al.}(1999)}]{SupernovaCosmologyProject:1998vns}
Perlmutter, S. {et~al.} 1999, Astrophys. J., 517, 565

\bibitem[{{Sandage}(1986)}]{bib:Sandage1986}
{Sandage}, A. 1986, \apj, 307, 1

\bibitem[{Tanoglidis {et~al.}(2015)Tanoglidis, Pavlidou, \& Tomaras}]{Tanoglidis:2014lea}
Tanoglidis, D., Pavlidou, V., \& Tomaras, T. 2015, JCAP, 12, 060

\bibitem[{Teerikorpi \& Chernin(2010)}]{Teerikorpi:2010zz}
Teerikorpi, P. \& Chernin, A.~D. 2010, Astron. Astrophys., 516, A93

\bibitem[{{Tully}(2015)}]{Tully:2015AJ}
{Tully}, R.~B. 2015, \aj, 149, 54

\bibitem[{Tully(2015)}]{Tully:2015opa}
Tully, R.~B. 2015, Astron. J., 149, 171

\bibitem[{Tully {et~al.}(2009)Tully, Rizzi, Shaya, Courtois, Makarov, \& Jacobs}]{Tully:2009EDD}
Tully, R.~B., Rizzi, L., Shaya, E.~J., {et~al.} 2009, Astron. J., 138, 323

\bibitem[{Tully {et~al.}(2023)}]{Tully:2022rbj}
Tully, R.~B. {et~al.} 2023, Astrophys. J., 944, 94

\bibitem[{{Wagner} \& {Benisty}(2025)}]{Wagner:2025wrp}
{Wagner}, J. \& {Benisty}, D. 2025, arXiv e-prints, arXiv:2501.13149

\bibitem[{Wagner {et~al.}(2025)Wagner, Benisty, \& Karachentsev}]{Wagner:2025wvh}
Wagner, J., Benisty, D., \& Karachentsev, I.~D. 2025 [\eprint[arXiv]{2510.24840}]

\end{thebibliography}

%\end{document}
%

\appendix

\onecolumn
\section{The Dataset}

\begin{longtable}[h!]{lrrrr|cccc|r}
\caption{\textit{Centaurus A and M83 group members and nearby galaxy, with their name, coordinates (J2000), distance with uncertainties, velocity from the Local group with uncertainty, references for data (1 :EDD ~\citep{Tully:2009EDD}, 2: \citep{Muller:2021,Muller:2025}) and usage in different calculations. The rows are sorted by Right Ascension. \label{tab:CenA-M83-group}}}\\   
\hline
\multicolumn{1}{c}{Name} &
\multicolumn{1}{c}{RA (deg)} &
\multicolumn{1}{c}{Dec (deg)} &
\multicolumn{1}{c}{Distance (Mpc)} &
\multicolumn{1}{c}{$v_\mathrm{LG}$ (km/s)} &
\multicolumn{4}{c}{Used in} &
\multicolumn{1}{c}{Ref} \\
\cline{6-9}
 & & & & &
\multicolumn{2}{c}{Hubble fit} & \multicolumn{2}{c}{Virial mass} & \\
\cline{6-7} \cline{8-9}
 & & & & &
Min & Maj & Min & Maj & \\
\hline
\endfirsthead
\hline
\multicolumn{10}{c}{{\bfseries \tablename\ \thetable{} -- continued from previous page}} \\
\hline
\multicolumn{1}{c}{Name} &
\multicolumn{1}{c}{RA (deg)} &
\multicolumn{1}{c}{Dec (deg)} &
\multicolumn{1}{c}{Distance (Mpc)} &
\multicolumn{1}{c}{$v_\mathrm{LG}$ (km/s)} &
\multicolumn{4}{c}{Used in} &
\multicolumn{1}{c}{Ref} \\
\cline{6-9}
 & & & & & \multicolumn{2}{c}{Hubble fit} & \multicolumn{2}{c}{Virial mass} & \\
\cline{6-7} \cline{8-9}
 & & & & & Min & Maj & Min & Maj & \\
\hline
\endhead
\hline
\endfoot
PGC2807061 & 105.1176 & -4.2071 & $6.52_{-0.26}^{+0.28}$ & $110.5 \pm 4$  &  &  &  &  &(1) \\
PGC020125 & 106.3238 & -58.5189 & $5.18_{-0.26}^{+0.27}$ & $278.7 \pm 4$  &  &  &  &  &(1) \\
ESO558-011 & 106.7376 & -22.0396 & $7.01_{-0.32}^{+0.33}$ & $493.0 \pm 3$  &  &  &  &  &(1) \\
UGC03755 & 108.4653 & 10.5211 & $7.52_{-0.17}^{+0.18}$ & $188.8 \pm 2$  &  &  &  &  &(1) \\
ESO059-001 & 112.8268 & -68.1875 & $4.53_{-0.06}^{+0.06}$ & $246.8 \pm 4$  & $\checkmark$ & $\checkmark$ &  &  &(1) \\
AGC174585 & 114.0429 & 9.9864 & $7.55_{-0.37}^{+0.39}$ & $216.9 \pm 8$  &  &  &  &  &(1) \\
UGC03974 & 115.4808 & 16.8025 & $7.98_{-0.07}^{+0.07}$ & $162.3 \pm 2$  &  &  &  &  &(1) \\
PGC021614 & 115.6320 & 16.5612 & $7.80_{-0.11}^{+0.11}$ & $172.8 \pm 4$  &  &  &  &  &(1) \\
PGC2807065 & 116.5250 & -28.4263 & $5.68_{-0.23}^{+0.24}$ & $217.7 \pm 7$  &  &  &  &  &(1) \\
AGC174605 & 117.5904 & 7.7944 & $9.38_{-0.17}^{+0.17}$ & $195.6 \pm 8$  &  &  &  &  &(1) \\
UGC04115 & 119.2590 & 14.3910 & $7.69_{-0.11}^{+0.11}$ & $222.3 \pm 6$  &  &  &  &  &(1) \\
ESO495-021 & 129.0633 & -26.4093 & $8.05_{-0.26}^{+0.26}$ & $571.9 \pm 9$  &  &  &  &  &(1) \\
ESO496-010 & 132.2749 & -26.3215 & $8.39_{-0.30}^{+0.32}$ & $519.4 \pm 10$  &  &  &  &  &(1) \\
PGC086668 & 135.7250 & 20.0747 & $8.63_{-0.27}^{+0.28}$ & $366.4 \pm 3$  &  &  &  &  &(1) \\
PGC2806961 & 137.2236 & 14.5829 & $9.38_{-0.46}^{+0.49}$ & $172.3 \pm 6$  &  &  &  &  &(1) \\
PGC1599237 & 138.4131 & 19.6188 & $8.13_{-0.33}^{+0.34}$ & $322.7 \pm 2$  &  &  &  &  &(1) \\
NGC2835 & 139.4703 & -22.3546 & $11.86_{-0.89}^{+0.97}$ & $602.1 \pm 2$  &  &  &  &  &(1) \\
PGC086669 & 139.8764 & 21.6036 & $9.08_{-0.49}^{+0.52}$ & $385.3 \pm 6$  &  &  &  &  &(1) \\
NGC2915 & 141.5493 & -76.6264 & $4.19_{-0.06}^{+0.06}$ & $189.0 \pm 3$  & $\checkmark$ & $\checkmark$ &  &  &(1) \\
NGC2903 & 143.0421 & 21.5014 & $8.95_{-0.36}^{+0.38}$ & $442.1 \pm 2$  &  &  &  &  &(1) \\
UGC05086 & 143.2034 & 21.4657 & $8.28_{-0.34}^{+0.35}$ & $393.9 \pm 34$  &  &  &  &  &(1) \\
ANTLIAB & 147.2337 & -25.9900 & $1.39_{-0.06}^{+0.07}$ & $82.2 \pm 6$  & $\checkmark$ & $\checkmark$ &  &  &(1) \\
UGC05288 & 147.8208 & 7.8283 & $11.27_{-0.66}^{+0.70}$ & $374.6 \pm 12$  &  &  &  &  &(1) \\
6dFJ0956376-092911 & 149.1567 & -9.4863 & $9.16_{-0.45}^{+0.48}$ & $372.6 \pm 41$  &  &  &  &  &(1) \\
UGC05373 & 150.0001 & 5.3322 & $1.40_{-0.02}^{+0.02}$ & $111.9 \pm 2$  & $\checkmark$ & $\checkmark$ &  &  &(1) \\
PGC029033 & 150.4092 & -8.2485 & $9.73_{-1.10}^{+1.24}$ & $202.2 \pm 3$  &  &  &  &  &(1) \\
NGC3109 & 150.7786 & -26.1595 & $1.31_{-0.05}^{+0.06}$ & $109.6 \pm 2$  & $\checkmark$ & $\checkmark$ &  &  &(1) \\
PGC029194 & 151.0165 & -27.3320 & $1.34_{-0.06}^{+0.06}$ & $65.3 \pm 3$  & $\checkmark$ & $\checkmark$ &  &  &(1) \\
UGC05456 & 151.8324 & 10.3620 & $10.28_{-0.99}^{+1.10}$ & $375.4 \pm 3$  &  &  &  &  &(1) \\
PGC029653 & 152.7533 & -4.6928 & $1.41_{-0.05}^{+0.05}$ & $94.0 \pm 1$  & $\checkmark$ & $\checkmark$ &  &  &(1) \\
UGC05797 & 159.8550 & 1.7185 & $10.19_{-0.98}^{+1.09}$ & $507.5 \pm 2$  &  &  &  &  &(1) \\
ESO376-016 & 160.8636 & -37.0439 & $6.76_{-0.30}^{+0.32}$ & $364.5 \pm 6$  & $\checkmark$ &  &  &  &(1) \\
NGC3351 & 160.9904 & 11.7035 & $9.73_{-0.13}^{+0.14}$ & $624.1 \pm 2$  &  &  &  &  &(1) \\
NGC3368 & 161.6904 & 11.8198 & $10.91_{-0.44}^{+0.46}$ & $739.3 \pm 2$  &  &  &  &  &(1) \\
PGC4689210 & 161.7216 & 12.7444 & $9.59_{-0.96}^{+1.07}$ & $491.7 \pm 5$  &  &  &  &  &(1) \\
PGC083339 & 161.7369 & 12.9992 & $9.42_{-0.55}^{+0.58}$ & $684.9 \pm 4$  &  &  &  &  &(1) \\
ESO318-013 & 161.9232 & -38.8537 & $6.76_{-0.12}^{+0.13}$ & $409.5 \pm 7$  & $\checkmark$ &  &  &  &(1) \\
NGC3377 & 161.9265 & 13.9856 & $10.19_{-0.09}^{+0.09}$ & $541.9 \pm 11$  &  &  &  &  &(1) \\
NGC3379 & 161.9566 & 12.5816 & $11.07_{-0.10}^{+0.10}$ & $759.2 \pm 6$  &  &  &  &  &(1) \\
NGC3384 & 162.0705 & 12.6294 & $9.20_{-0.08}^{+0.09}$ & $414.5 \pm 6$  &  &  &  &  &(1) \\
AGC205315 & 162.4683 & 12.5392 & $10.00_{-0.49}^{+0.52}$ & $630.5 \pm 8$  &  &  &  &  &(1) \\
ESO215-009 & 164.3750 & -48.1839 & $5.35_{-0.07}^{+0.07}$ & $288.8 \pm 2$  & $\checkmark$ &  &  &  &(1) \\
NGC3621 & 169.5687 & -32.8140 & $6.82_{-0.28}^{+0.29}$ & $438.1 \pm 3$  & $\checkmark$ & $\checkmark$ &  &  &(1) \\
NGC3627 & 170.0625 & 12.9910 & $11.02_{-0.50}^{+0.52}$ & $584.5 \pm 2$  &  &  &  &  &(1) \\
PGC683190 & 173.2957 & -32.9624 & $5.47_{-0.15}^{+0.15}$ & $419.0 \pm 9$  & $\checkmark$ & $\checkmark$ & $\checkmark$ & $\checkmark$ &(1) \\
ESO320-014 & 174.4716 & -39.2206 & $5.86_{-0.11}^{+0.11}$ & $362.2 \pm 4$  & $\checkmark$ & $\checkmark$ &  &  &(1) \\
ESO379-007 & 178.6798 & -33.5588 & $5.32_{-0.17}^{+0.17}$ & $364.6 \pm 3$  & $\checkmark$ & $\checkmark$ & $\checkmark$ & $\checkmark$ &(1) \\
ESO572-034 & 179.7420 & -19.0299 & $10.28_{-0.37}^{+0.39}$ & $871.5 \pm 4$  &  &  &  &  &(1) \\
ESO379-024 & 181.2362 & -35.7431 & $5.35_{-0.24}^{+0.25}$ & $355.5 \pm 6$  & $\checkmark$ & $\checkmark$ & $\checkmark$ & $\checkmark$ &(1) \\
ESO321-014 & 183.4571 & -38.2310 & $3.25_{-0.15}^{+0.15}$ & $334.8 \pm 3$  &  &  & $\checkmark$ & $\checkmark$ &(1) \\
IC3104 & 184.6920 & -79.7259 & $2.30_{-0.02}^{+0.02}$ & $170.2 \pm 4$  & $\checkmark$ & $\checkmark$ & $\checkmark$ & $\checkmark$ &(1) \\
NGC4298 & 185.3867 & 14.6061 & $14.52_{-1.28}^{+1.40}$ & $1029.6 \pm 5$  &  &  &  &  &(1) \\
PGC040045 & 185.5317 & 15.7992 & $9.08_{-1.02}^{+1.15}$ & $1205.6 \pm 2$  &  &  &  &  &(1) \\
NGC4328 & 185.8333 & 15.8205 & $14.52_{-1.52}^{+1.70}$ & $386.3 \pm 10$  &  &  &  &  &(1) \\
UGC07512 & 186.4212 & 2.1593 & $11.59_{-0.42}^{+0.43}$ & $1353.0 \pm 2$  &  &  &  &  &(1) \\
NGC4413 & 186.6342 & 12.6106 & $14.72_{-0.27}^{+0.27}$ & $-10.6 \pm 4$  &  &  &  &  &(1) \\
PGC086635 & 187.0206 & 22.2908 & $6.17_{-0.25}^{+0.26}$ & $543.9 \pm 1$  & $\checkmark$ & $\checkmark$ & $\checkmark$ & $\checkmark$ &(1) \\
NGC4455 & 187.1837 & 22.8215 & $6.46_{-0.26}^{+0.27}$ & $586.8 \pm 3$  & $\checkmark$ & $\checkmark$ & $\checkmark$ & $\checkmark$ &(1) \\
PGC041395 & 187.7652 & 1.6758 & $9.20_{-0.57}^{+0.61}$ & $954.3 \pm 4$  &  &  &  &  &(1) \\
AGC229385 & 188.0429 & 20.4233 & $10.62_{-0.93}^{+1.02}$ & $1282.3 \pm 4$  &  &  &  &  &(1) \\
NGC4437 & 188.1895 & 0.1146 & $8.17_{-0.44}^{+0.46}$ & $972.0 \pm 2$  &  &  &  &  &(1) \\
PGC042120 & 189.3085 & -10.4979 & $9.59_{-0.30}^{+0.31}$ & $562.5 \pm 8$  & $\checkmark$ & $\checkmark$ &  &  &(1) \\
PGC3097708 & 189.3995 & -8.8671 & $9.55_{-0.43}^{+0.45}$ & $917.5 \pm 5$  &  &  &  &  &(1) \\
NGC4592 & 189.8281 & -0.5320 & $8.67_{-0.35}^{+0.37}$ & $913.5 \pm 2$  &  &  &  &  &(1) \\
NGC4594 & 189.9976 & -11.6230 & $9.12_{-0.33}^{+0.34}$ & $891.3 \pm 5$  &  &  &  &  &(1) \\
NGC4600 & 190.0958 & 3.1177 & $9.08_{-0.29}^{+0.30}$ & $713.3 \pm 30$  &  & $\checkmark$ &  &  &(1) \\
ESO381-018 & 191.1772 & -35.9666 & $5.32_{-0.12}^{+0.12}$ & $362.7 \pm 8$  & $\checkmark$ & $\checkmark$ & $\checkmark$ & $\checkmark$ &(1) \\
ESO381-020 & 191.5017 & -33.8381 & $5.35_{-0.12}^{+0.12}$ & $335.4 \pm 3$  & $\checkmark$ & $\checkmark$ & $\checkmark$ & $\checkmark$ &(1) \\
PGC043072 & 191.5638 & 10.2055 & $9.42_{-0.91}^{+1.00}$ & $1031.9 \pm 2$  &  &  &  &  &(1) \\
HIPASSJ1247-77 & 191.8908 & -77.5826 & $3.39_{-0.26}^{+0.28}$ & $153.8 \pm 4$  &  &  & $\checkmark$ & $\checkmark$ &(1) \\
SDSSJ124936.90+215505.7 & 192.4035 & 21.9178 & $6.34_{-0.29}^{+0.30}$ & $491.1 \pm 4$  & $\checkmark$ & $\checkmark$ & $\checkmark$ & $\checkmark$ &(1) \\
PGC086644 & 192.9423 & 21.7357 & $5.83_{-0.21}^{+0.22}$ & $534.6 \pm 4$  & $\checkmark$ & $\checkmark$ & $\checkmark$ & $\checkmark$ &(1) \\
PGC043851 & 193.4868 & -12.1059 & $8.95_{-0.56}^{+0.60}$ & $634.8 \pm 2$  &  & $\checkmark$ &  &  &(1) \\
ESO443-009 & 193.7232 & -28.3408 & $5.83_{-0.29}^{+0.30}$ & $407.6 \pm 4$  & $\checkmark$ & $\checkmark$ & $\checkmark$ & $\checkmark$ &(1) \\
PGC044055 & 193.9185 & 19.2092 & $4.72_{-0.15}^{+0.15}$ & $362.3 \pm 4$  &  &  & $\checkmark$ & $\checkmark$ &(1) \\
NGC4826 & 194.1825 & 21.6820 & $4.31_{-0.06}^{+0.06}$ & $363.5 \pm 4$  &  &  & $\checkmark$ & $\checkmark$ &(1) \\
UGC08091 & 194.6678 & 14.2173 & $2.14_{-0.11}^{+0.11}$ & $138.4 \pm 2$  & $\checkmark$ & $\checkmark$ & $\checkmark$ & $\checkmark$ &(1) \\
PGC044681 & 194.9857 & -19.4114 & $7.11_{-0.29}^{+0.30}$ & $619.1 \pm 2$  & $\checkmark$ & $\checkmark$ &  &  &(1) \\
PGC044982 & 195.5598 & -17.2378 & $5.62_{-0.18}^{+0.18}$ & $549.5 \pm 2$  & $\checkmark$ & $\checkmark$ & $\checkmark$ & $\checkmark$ &(1) \\
PGC045084 & 195.8199 & -17.4229 & $5.89_{-0.24}^{+0.25}$ & $546.6 \pm 2$  & $\checkmark$ & $\checkmark$ & $\checkmark$ & $\checkmark$ &(1) \\
ESO269-037 & 195.8875 & -46.5842 & $3.15_{-0.09}^{+0.09}$ & $481.4 \pm 2$  & $\checkmark$ & $\checkmark$ & $\checkmark$ & $\checkmark$ &(2) \\
ESO269-037 & 195.8883 & -46.5869 & $3.08_{-0.10}^{+0.10}$ & $481.4 \pm 6$  & $\checkmark$ & $\checkmark$ & $\checkmark$ & $\checkmark$ &(1) \\
PGC166152 & 196.2597 & -40.0823 & $5.81_{-0.34}^{+0.36}$ & $361.9 \pm 2$  & $\checkmark$ & $\checkmark$ & $\checkmark$ & $\checkmark$ &(1) \\
NGC4945 & 196.3638 & -49.4679 & $3.39_{-0.05}^{+0.05}$ & $299.6 \pm 3$  &  &  & $\checkmark$ & $\checkmark$ &(1) \\
ESO269-058 & 197.6333 & -46.9908 & $3.75_{-0.02}^{+0.02}$ & $140.5 \pm 18$  &  &  & $\checkmark$ &  &(2) \\
KK189 & 198.1883 & -41.8320 & $4.21_{-0.17}^{+0.17}$ & $501.5 \pm 4$  &  &  & $\checkmark$ & $\checkmark$ &(2) \\
ESO269-068 & 198.2985 & -43.2649 & $3.65_{-0.03}^{+0.03}$ & $393.6 \pm 96$  &  &  & $\checkmark$ & $\checkmark$ &(1) \\
NGC5068 & 199.7295 & -21.0392 & $5.04_{-0.20}^{+0.21}$ & $471.3 \pm 2$  &  &  & $\checkmark$ & $\checkmark$ &(1) \\
PGC166163 & 200.2845 & -31.5291 & $5.50_{-0.25}^{+0.26}$ & $344.6 \pm 3$  & $\checkmark$ & $\checkmark$ & $\checkmark$ & $\checkmark$ &(1) \\
KKs54 & 200.3829 & -31.8864 & $3.75_{-0.10}^{+0.10}$ & $394.0 \pm 11$  &  &  & $\checkmark$ & $\checkmark$ &(2) \\
PGC046663 & 200.4461 & -45.0624 & $3.87_{-0.11}^{+0.11}$ & $489.7 \pm 15$  &  &  & $\checkmark$ & $\checkmark$ &(1) \\
NGC5102 & 200.4875 & -36.6297 & $3.74_{-0.39}^{+0.39}$ & $227.0 \pm 18$  &  &  & $\checkmark$ &  &(2) \\
KK197 & 200.5086 & -42.5359 & $3.84_{-0.04}^{+0.04}$ & $395.6 \pm 3$  &  &  & $\checkmark$ & $\checkmark$ &(2) \\
KKs55 & 200.5500 & -42.7311 & $3.85_{-0.07}^{+0.07}$ & $282.3 \pm 14$  &  &  & $\checkmark$ & $\checkmark$ &(2) \\
dw1322-39 & 200.6336 & -39.9060 & $2.95_{-0.05}^{+0.05}$ & $413.2 \pm 10$  &  &  & $\checkmark$ & $\checkmark$ &(2) \\
dw1323-40b & 200.9809 & -40.8361 & $3.91_{-0.61}^{+0.61}$ & $253.4 \pm 12$  &  &  & $\checkmark$ & $\checkmark$ &(2) \\
PGC046885 & 201.1501 & -30.9719 & $4.66_{-0.13}^{+0.13}$ & $267.2 \pm 3$  &  &  & $\checkmark$ & $\checkmark$ &(1) \\
dw1323-40a & 201.2233 & -40.7612 & $3.73_{-0.15}^{+0.15}$ & $207.1 \pm 14$  &  &  & $\checkmark$ & $\checkmark$ &(2) \\
NGC5128/CenA & 201.3672 & -43.0181 & $3.60_{-0.07}^{+0.07}$ & $305.7 \pm 6$  &  &  &  &  &(1) \\
IC4247 & 201.6852 & -30.3625 & $5.06_{-0.09}^{+0.09}$ & $200.1 \pm 8$  & $\checkmark$ & $\checkmark$ & $\checkmark$ & $\checkmark$ &(1) \\
KK203 & 201.8681 & -45.3524 & $3.78_{-0.25}^{+0.25}$ & $57.4 \pm 10$  &  &  &  &  &(2) \\
ESO324-024 & 201.9042 & -41.4806 & $3.78_{-0.09}^{+0.09}$ & $271.5 \pm 18$  &  &  & $\checkmark$ & $\checkmark$ &(2) \\
NGC5206 & 203.4292 & -48.1511 & $3.21_{-0.01}^{+0.01}$ & $334.1 \pm 6$  &  &  & $\checkmark$ & $\checkmark$ &(2) \\
ESO270-017 & 203.6951 & -45.5473 & $6.79_{-0.15}^{+0.16}$ & $584.2 \pm 4$  & $\checkmark$ & $\checkmark$ & $\checkmark$ & $\checkmark$ &(1) \\
ESO444-078 & 204.1284 & -29.2354 & $5.30_{-0.07}^{+0.07}$ & $361.6 \pm 2$  & $\checkmark$ &  & $\checkmark$ & $\checkmark$ &(1) \\
PGC677373 & 204.2511 & -33.3625 & $4.45_{-0.30}^{+0.32}$ & $370.7 \pm 4$  &  &  & $\checkmark$ & $\checkmark$ &(1) \\
NGC5236/M83 & 204.2536 & -29.8656 & $4.79_{-0.09}^{+0.09}$ & $296.3 \pm 2$  &  &  &  &  &(1) \\
PGC592761 & 204.3561 & -39.8887 & $4.97_{-0.25}^{+0.26}$ & $257.2 \pm 3$  &  &  & $\checkmark$ & $\checkmark$ &(1) \\
NGC5237 & 204.4083 & -42.8475 & $3.33_{-0.02}^{+0.02}$ & $122.0 \pm 4$  &  &  & $\checkmark$ & $\checkmark$ &(2) \\
NGC5253 & 204.9792 & -31.6400 & $3.55_{-0.03}^{+0.03}$ & $192.8 \pm 3$  &  &  & $\checkmark$ & $\checkmark$ &(2) \\
IC4316 & 205.0761 & -28.8942 & $4.25_{-0.08}^{+0.08}$ & $371.1 \pm 3$  &  &  & $\checkmark$ & $\checkmark$ &(1) \\
NGC5264 & 205.4031 & -29.9127 & $4.68_{-0.15}^{+0.15}$ & $269.4 \pm 2$  &  &  & $\checkmark$ & $\checkmark$ &(1) \\
dw1341-43 & 205.4032 & -43.8553 & $3.53_{-0.04}^{+0.04}$ & $397.7 \pm 14$  &  &  & $\checkmark$ & $\checkmark$ &(2) \\
KKs57 & 205.4079 & -42.5797 & $3.84_{-0.47}^{+0.47}$ & $274.8 \pm 17$  &  &  & $\checkmark$ &  &(2) \\
KK211 & 205.5208 & -45.2050 & $3.68_{-0.14}^{+0.14}$ & $359.8 \pm 31$  &  &  & $\checkmark$ & $\checkmark$ &(2) \\
dw1342-43 & 205.6837 & -43.2548 & $2.90_{-0.14}^{+0.14}$ & $273.3 \pm 8$  &  &  & $\checkmark$ & $\checkmark$ &(2) \\
ESO325-011 & 206.2542 & -41.8597 & $3.33_{-0.06}^{+0.06}$ & $313.1 \pm 3$  &  &  & $\checkmark$ & $\checkmark$ &(1) \\
KKs58 & 206.5031 & -36.3289 & $3.36_{-0.10}^{+0.10}$ & $255.9 \pm 5$  &  &  & $\checkmark$ & $\checkmark$ &(2) \\
HIPASSJ1348-37 & 207.1396 & -37.9681 & $5.52_{-0.18}^{+0.18}$ & $359.0 \pm 6$  & $\checkmark$ & $\checkmark$ & $\checkmark$ & $\checkmark$ &(1) \\
KK221 & 207.1917 & -46.9974 & $3.82_{-0.07}^{+0.07}$ & $267.8 \pm 13$  &  &  & $\checkmark$ &  &(2) \\
ESO383-087 & 207.3250 & -36.0614 & $3.19_{-0.03}^{+0.03}$ & $107.7 \pm 2$  &  &  & $\checkmark$ & $\checkmark$ &(2) \\
PGC3097113 & 207.8286 & -46.9735 & $5.60_{-0.30}^{+0.32}$ & $291.3 \pm 5$  & $\checkmark$ & $\checkmark$ & $\checkmark$ & $\checkmark$ &(1) \\
PGC2807150 & 208.6396 & 4.2444 & $2.55_{-0.29}^{+0.32}$ & $206.5 \pm 3$  &  & $\checkmark$ & $\checkmark$ & $\checkmark$ &(1) \\
ESO384-016 & 209.2559 & -35.3329 & $4.39_{-0.06}^{+0.06}$ & $293.6 \pm 9$  &  &  & $\checkmark$ & $\checkmark$ &(1) \\
NGC5398 & 210.3400 & -33.0637 & $12.13_{-1.07}^{+1.17}$ & $1016.2 \pm 3$  &  &  &  &  &(1) \\
NGC5408 & 210.8383 & -41.3766 & $5.20_{-0.16}^{+0.17}$ & $286.5 \pm 3$  & $\checkmark$ & $\checkmark$ & $\checkmark$ & $\checkmark$ &(1) \\
DDO187 & 213.9854 & 23.0556 & $2.25_{-0.04}^{+0.04}$ & $171.7 \pm 2$  & $\checkmark$ & $\checkmark$ & $\checkmark$ & $\checkmark$ &(1) \\
PGC051659 & 217.0149 & -46.3052 & $3.53_{-0.35}^{+0.39}$ & $177.1 \pm 2$  &  &  & $\checkmark$ & $\checkmark$ &(1) \\
NGC5643 & 218.1696 & -44.1745 & $12.13_{-0.60}^{+0.63}$ & $981.3 \pm 14$  &  &  &  &  &(1) \\
ESO222-010 & 218.7626 & -49.4217 & $3.08_{-0.12}^{+0.13}$ & $404.3 \pm 4$  &  & $\checkmark$ & $\checkmark$ & $\checkmark$ &(1) \\
ESO272-025 & 220.8561 & -44.7051 & $3.82_{-0.16}^{+0.16}$ & $427.0 \pm 3$  &  &  & $\checkmark$ &  &(1) \\
ESO223-009 & 225.2858 & -48.2918 & $6.17_{-0.06}^{+0.06}$ & $389.1 \pm 2$  & $\checkmark$ & $\checkmark$ &  &  &(1) \\
ESO274-001 & 228.5559 & -46.8091 & $2.73_{-0.10}^{+0.10}$ & $336.7 \pm 3$  & $\checkmark$ & $\checkmark$ & $\checkmark$ & $\checkmark$ &(1) \\
ESO137-018 & 245.2464 & -60.4876 & $5.73_{-0.21}^{+0.21}$ & $421.2 \pm 5$  & $\checkmark$ &  &  &  &(1) \\
IC4662 & 266.7892 & -64.6391 & $2.49_{-0.02}^{+0.02}$ & $143.0 \pm 3$  & $\checkmark$ & $\checkmark$ &  &  &(1) \\
IC4710 & 277.1581 & -66.9822 & $7.21_{-0.20}^{+0.20}$ & $580.1 \pm 4$  &  &  &  &  &(1) \\
ESO104-022 & 283.9215 & -64.8114 & $8.75_{-0.24}^{+0.25}$ & $653.2 \pm 4$  &  &  &  &  &(1) \\
NGC6744 & 287.4413 & -63.8577 & $9.16_{-0.41}^{+0.43}$ & $716.9 \pm 16$  &  &  &  &  &(1) \\
ESO104-044 & 287.8461 & -64.2190 & $9.51_{-0.76}^{+0.82}$ & $639.7 \pm 8$  &  &  &  &  &(1) \\
ESO594-004 & 292.4959 & -17.6790 & $1.06_{-0.04}^{+0.04}$ & $21.0 \pm 3$  & $\checkmark$ & $\checkmark$ &  &  &(1) \\
IC4870 & 294.4065 & -65.8119 & $8.32_{-0.19}^{+0.19}$ & $739.4 \pm 3$  &  &  &  &  &(1) \\
ALFAZOAJ1952+1428 & 298.0491 & 14.4733 & $8.20_{-0.51}^{+0.55}$ & $520.8 \pm 6$  &  &  &  &  &(1) \\
ESO461-036 & 300.9889 & -31.6816 & $6.89_{-0.28}^{+0.29}$ & $475.0 \pm 6$  &  &  &  &  &(1) \\
IC4951 & 302.3822 & -61.8506 & $8.99_{-0.56}^{+0.60}$ & $701.2 \pm 3$  &  &  &  &  &(1) \\
IC5052 & 313.0212 & -69.2014 & $5.37_{-0.15}^{+0.15}$ & $445.1 \pm 3$  &  &  &  &  &(1) \\
NGC7090 & 324.1205 & -54.5576 & $9.29_{-0.25}^{+0.26}$ & $784.8 \pm 4$  &  &  &  &  &(1) \\
IC5152 & 330.6723 & -51.2966 & $1.91_{-0.03}^{+0.04}$ & $75.0 \pm 2$  & $\checkmark$ & $\checkmark$ &  &  &(1) \\
ESO238-005 & 335.6270 & -48.4040 & $7.83_{-0.18}^{+0.18}$ & $671.1 \pm 3$  &  &  &  &  &(1) \\
Tucana & 340.4580 & -64.4206 & $0.90_{-0.02}^{+0.02}$ & $72.7 \pm 4$  & $\checkmark$ & $\checkmark$ &  &  &(1) \\
AF7448 001 & 344.8981 & 16.7654 & $7.80_{-0.65}^{+0.71}$ & $619.8 \pm 6$  &  &  &  &  &(1) \\
ESO407-018 & 351.6160 & -32.3886 & $2.17_{-0.08}^{+0.08}$ & $98.8 \pm 3$  & $\checkmark$ & $\checkmark$ &  &  &(1) \\
ESO347-017 & 351.7344 & -37.3469 & $8.24_{-0.30}^{+0.31}$ & $701.6 \pm 3$  &  &  &  &  &(1) \\
IC5332 & 353.6145 & -36.1016 & $8.79_{-0.40}^{+0.41}$ & $719.5 \pm 3$  &  &  &  &  &(1) \\
ESO471-006 & 355.9379 & -31.9593 & $4.27_{-0.15}^{+0.16}$ & $301.5 \pm 9$  &  &  &  &  &(1) \\
ESO149-003 & 358.0122 & -52.5774 & $6.85_{-0.16}^{+0.16}$ & $513.1 \pm 5$  &  &  &  &  &(1) \\
NGC7793 & 359.4573 & -32.5910 & $3.77_{-0.10}^{+0.11}$ & $251.3 \pm 3$  &  &  &  &  &(1) \\
PGC704814 & 359.6696 & -31.4676 & $3.58_{-0.13}^{+0.13}$ & $298.7 \pm 89$  &  &  &  &  &(1) \\
\end{longtable}

\end{document}